\documentclass{emulateapj}

\usepackage{color}

\def\be{\begin{equation}}
\def\ee{\end{equation}}
\def\nms{\mathsurround=0pt}
\def\oversim#1#2{\lower 4pt\vbox{\baselineskip 0pt \lineskip 1pt
    \ialign{$\nms#1\hfil##\hfil$\crcr#2\crcr\sim\crcr}}}
\def\ga{\mathrel{\mathpalette\oversim>}}
\def\la{\mathrel{\mathpalette\oversim<}}
\def\arcdeg{{^{\circ}}}

\def\bh{M_{\bullet}}
\def\bhsp{M_{\bullet,1}}
\def\bhs{M_{\bullet,2}}

\def\pc{\rm pc}
\def\kpc{\rm ~kpc}
\def\msun{M_{\odot}}

\def\AU{\rm ~AU}
\def\kms{\rm ~km~s^{-1}}

\def\mpc{\rm mpc}

\begin{document}
\shorttitle{Spatial and velocity distributions of HVSs}
\shortauthors{Zhang, Lu, \& Yu}

\title{The spatial and velocity distributions of hypervelocity stars }
\author{Fupeng Zhang$^1$, Youjun Lu$^1$, and Qingjuan Yu$^{2}$}
\affil{$^1$~National Astronomical Observatories, Chinese Academy of Sciences,
Beijing, 100012, China; luyj@nao.cas.cn \\
$^2$~Kavli Institute for Astronomy and Astrophysics,
Peking University, Beijing, 100871, China; yuqj@kiaa.pku.edu.cn
}

\begin{abstract}

Hypervelocity stars (HVSs) found in the Galactic halo are believed to be the
dynamical products of interactions between (binary) stars and the massive
black hole(s) (MBH) in the Galactic center (GC).  In this paper, we investigate
how the spatial and velocity distributions of HVSs are connected with their
originations, ejecting mechanisms, and the dynamical environment in the GC. It has
been shown that the detected HVSs are spatially consistent with being located
on two thin disks (Lu et al.), one of which has the same orientation as
the clockwise rotating stellar disk in the GC. Here, we perform a large number
of three-body
experiments of the interactions between the MBH and binary stars bound to it; and
we find that the probability of ejecting HVSs is substantially enhanced by
multiple encounters between the MBH and binary stars at a distance
substantially larger than their initial tidal breakup radii. Assuming that
the HVS progenitors are originated from the two thin disks,
our simulations show that the distributions of
the HVS inclination relative to the disk planes can be well reproduced by either the
mechanism of tidal breakup of binary stars or the mechanism of ejecting HVSs
by a hypothetical binary black hole (BBH) in the GC. However, an isotropical origination of HVS
progenitors is inconsistent with the observed inclination distribution.
Assuming that the detected HVSs were ejected out by tidal breakup
of binary stars, its velocity distribution can be reproduced if 
their progenitors diffuse onto low angular momentum orbits
slowly and most of the progenitors were broken up at relatively large distances
because of multiple encounters. Assuming that the HVSs were ejected out by a BBH
within the allowed parameter space in the GC, our simulations produce
relatively flatter spectra at the high-velocity end compared to the observed
ones; however, the BBH mechanism cannot be statistically ruled out, yet.
Future deep surveys of HVSs and better statistics of the HVS spatial and
velocity distributions should enable to distinguish the ejection mechanisms of
HVSs and shed new light on the dynamical environment surrounding the central
MBH.

\end{abstract}


\section{Introduction}\label{sec:intro}

Hypervelocity stars (HVSs) discovered recently are mainly B type stars with
mass $\sim 3$-$4\msun$ and radial velocities up to $700 \kms$ escaping away from
the Galactic halo \citep{Brown05,Hirsch05,Edelmann05,Brown07, Brown09a,
Bromley09}. These HVSs are believed to be ejected out of the Galactic center
(GC) by dynamical interactions of (binary) stars with the central massive black
hole(s) \citep[MBH;][]{Brown09a,Brown09b}. There are mainly three mechanisms that
can produce such high-velocity stars: (1) the tidal breakup of binary stars in
the immediate vicinity of the central MBH \citep[hereafter TBK
mechanism;][]{Hills88,YT03,Bromley06,Sari10}; (2) three-body interactions of
single stars with a hypothetical low-mass ratio binary black holes (BBHs) in the
GC \citep[hereafter, BBH mechanism;][]{YT03, Gualandris05, Levin06, Baumgardt06,
Sesana06, LB08}; and (3) interactions of single stars with a cluster of stellar
mass black holes in the vicinity of the central MBH \citep[hereafter, SBHs
mechanism;][]{OL08}.\footnote{ Alternative models, other than the GC origin,
are also proposed to explain the HVSs by \citet{Heber08}, \citet{Abadi09},
\citet{TF09}, and \citet{WH09}.}

To determine which mechanism should be responsible for the detected HVSs is an
important issue in understanding the stellar dynamics around the central MBH.
Some observational properties of HVSs, such as the binarity, rotational
velocity, metallicity, ejection rate/observed frequency, spatial and velocity
distribution, etc., are proposed to be helpful in distinguishing these
mechanisms \citep[e.g.,][]{Luetal07,H07,Sesana07,Przybilla08, LMB08, Perets09a,
Perets09b}. However, one uncertain factor in these mechanisms is where 
the initial progenitors come from. Different origins of the HVS progenitors 
may also result in different distributions of HVS properties, especially their
spatial distribution and the velocity distribution. The main aim of this paper
is to address this problem through Monte Carlo numerical simulations under the
constraints from current observations.

The spatial distribution of the detected HVSs is compatible with both the TBK
and the BBH mechanism if the progenitors of these HVSs are originated from
disk-like structures in the vicinity of the central MBH (e.g., the clockwise
rotating young stellar (CWS) disk in the GC).  The preliminary results that
demonstrate the disk(s) origination of the HVS progenitors have been summarized
in \citet*{Luetal10}. We shall further demonstrate in this paper that an
isotropic distribution of HVS progenitors is inconsistent with the spatial
distribution of the detected HVSs if they are originated from the GC. However,
the disk(s) origination of the detected HVSs is consistent with the
distribution of the inclination angles (relative to the disk(s)) of the
detected HVSs which further strengthen the conclusions made in
\citet*{Luetal10}. 

The velocity distribution of HVSs is related to not only the production
mechanism but also the origin of their progenitors. \citet{Sesana07} have
studied the velocity distribution of HVSs. They found that the velocity
distribution of HVSs produced by the TBK mechanism for unbound injecting stellar
binaries seems to be consistent with the then detected HVSs though with limited
statistics, while the HVS velocity distribution produced by the BBH mechanism
appears to be too flat in comparison with the observations. Their results
suggest that the HVS velocity distribution may be useful in distinguishing
the ejection mechanisms.  In this paper, we shall further investigate the
effects on the velocity distribution of HVSs due to different origins of the
HVS progenitors, e.g., those (binary) stars initially unbound to the MBH but
later injected into the immediate vicinity of the MBH due to some unknown
perturbations, and those (binary) stars initially bound to the MBH but later
evolved onto highly eccentric orbits and migrated into the immediate vicinity
of the MBH.

The paper is organized as follows. In Section~\ref{sec:obs}, we first summarize
the observational results on the spatial and velocity distribution of the
detected HVSs. In Section~\ref{sec:TBK}, we explore the detailed dynamics of
interactions between binary stars on bound orbits and a central MBH. The
consequences of these interactions are different from that between the unbound
binary stars on parabolic (or hyperbolic) orbits and the MBH intensively
investigated in the literature \citep[e.g.,][]{Hills88,Bromley06,Sesana07}. The
reason is that the stellar binary may experience multiple close encounters
with the MBH in the former case, while it only experiences a single close
encounter in the latter case.  Assuming realistic distributions of the
properties of the initial stellar binaries, we then simulate both the spatial
distribution and the velocity distribution of HVSs produced by the TBK
mechanism and compare the numerical results with the observations in
Section~\ref{sec:Result}.\footnote{The cases of interactions between unbound
binaries and the central MBH have been extensively studied by
\citet{Bromley06}.} In Section~\ref{sec:BBH}, we also explore the interactions
between single stars on bound orbits with a hypothesized BBH in the GC. These
single stars are assumed to be injected into the immediate vicinity of the BBH
from disk-like stellar structures (e.g., the CWS disk) surrounding the BBH,
which is different from that adopted in \citet{Sesana07}.  With reasonable but
simple assumptions on the parameters of the hypothetical BBH, the spatial and
velocity distributions of the ejected HVSs are obtained. Comparison between the
simulation results and the observations are also discussed in
Section~\ref{sec:BBH}. The conclusions are given in
Section~\ref{sec:Conclusion}.

\section{Spatial and velocity distributions of the detected HVSs}\label{sec:obs}

Surveys of HVSs have detected 16 HVSs unbound to the Galactic halo, 8 bound
HVSs, and 4 HVS candidates
\citep{Brown05,Hirsch05,Edelmann05,Brown07,Brown09a}. We summarize their
spatial and velocity distributions in this section.

\subsection{Spatial distribution}\label{subsec:sd}

\begin{figure*}
\centerline{\includegraphics[width=4.5in]{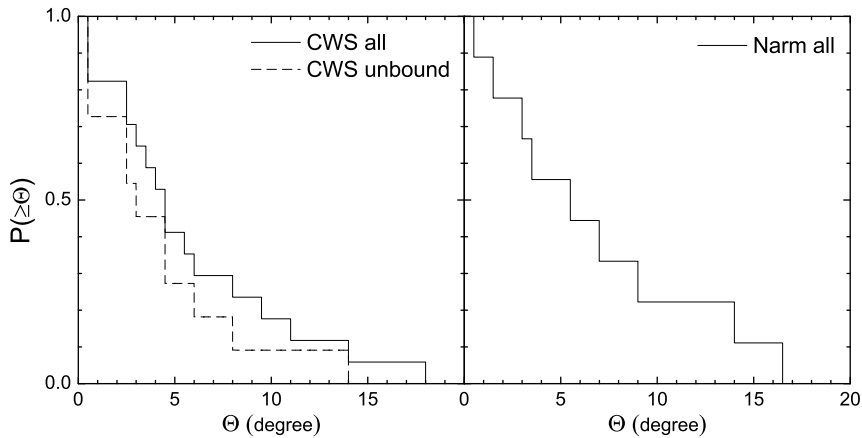}}
\caption{ 
Observational inclination distributions of HVSs relative to their best-fit
planes ($\Theta$CDF, $P(\geq \Theta)$).  The left panel is for the first population
of HVSs which are spatially associated with a disk plane with an orientation
similar to the CWS disk, and the right panel for the second population of the
HVSs which are spatially associated with a disk plane with an orientation
similar to the northern arm (Narm) of the minispiral or the warped outer part
of the CWS disk. The solid histograms represent the distributions of all the
detected HVSs, for each corresponding population,
including the unbound HVSs, the bound HVSs and the HVS candidates. The distribution
of the unbound HVSs is also
shown for the first population by a dashed line. See Section~\ref{subsec:sd}.
}
\label{fig:f1}
\end{figure*}

The spatial distribution of the HVSs detected so far is probably anisotropic
\citep{Abadi09,Brown09b}. \citet{Luetal10} use great circles to fit the spatial
distribution of the detected HVSs projected on the sky of an observer located
at the GC, and they find that the distribution can be best fitted by two great
circles.  Their results suggest that the spatial distribution of the detected HVSs is consistent with being located on the planes
of two thin disks \citep*{Luetal10}: (1) eleven of the unbound HVSs (plus four
bound ones and two candidates; totally 17 objects) are spatially associated to a thin disk plane
with an orientation almost the same as that of 
CWS disk located within half a parsec from the central MBH (see
\citealt{LB03,LuJ09,Paumard06,Bartko09a,Bartko09b}); (2) four of the
unbound HVSs (plus three bound ones and two candidates; totally 9 objects) are spatially
associated to a thin disk plane with an orientation similar to that of the
northern arm of the minispiral (or also the outer warped part of the CWS disk)
in the GC. The normals of the best-fit disk planes for these two HVS
populations are ($l$,$b$)$=$($311\arcdeg$, $-14\arcdeg$) and ($176\arcdeg$, $-53\arcdeg$)
in Galactic coordinates, respectively \citep{Luetal10}. Hereafter, we refer to those 
detected HVSs associated with the above two best-fit planes as the first population and the second population of HVSs, respectively. We denote the
inclination angle of each HVS to its corresponding best-fit plane by $\Theta$
and describe the spatial distribution of the HVSs by a normalized cumulative
distribution function of their inclination angles $P(\geq\Theta)$
(hereafter, $\Theta$CDF), which represents the number fraction of the HVSs with
inclination angles higher than $\Theta$. The observational $\Theta$CDFs for
both populations of the HVSs are shown in Figure~\ref{fig:f1} and will be
compared with the distributions obtained from numerical models in Sections
\ref{sec:Result} and \ref{sec:BBH}.  For each population, we shall compare the
$P(\geq\Theta)$ of all the HVSs (including unbound HVSs, bound HVSs, and HVS
candidates) instead of only unbound ones, because (1) for the first HVS
population, our Kolmogorov$-$Smirnov (K-S) test finds a likelihood of $0.94$ that
the unbound HVSs and all the HVSs are drawn from the same $\Theta$CDF; (2) for
the second population, the number of the unbound HVSs is only 4 and the error
due to Poisson noise in the $\Theta$CDF is substantial, therefore we do not show their $\Theta$CDF in
Figure~\ref{fig:f1} (and $v$CDF in Figure~\ref{fig:f2} below, either).

The gravitational potential of the Galaxy is not exactly spherical, and its
non-spherical component may deflect the radial trajectories of HVSs after they
were ejected from the GC \citep[e.g.,][]{YM07}. Given the distance and the
velocity span ($30\kpc<R<130\kpc$ and $690 \kms<v<980 \kms$, see
Section~\ref{subsec:vd}) of the detected HVSs, however, the deviation due to
the flattening of the Galactic disk and the triaxiality of the Galactic halo is
at most several degrees (typically $2\arcdeg$) as demonstrated in \citet{YM07}.
Compared with the standard deviation of $\Theta$ ($\sim7\arcdeg-8\arcdeg$)
shown in Figure~\ref{fig:f1}, that small deviation can be ignored.

\subsection{Velocity distribution}\label{subsec:vd}

The detected HVSs have been decelerated in the Galactic
gravitational potential after they were ejected from the GC. To easily compare
with results obtained from theoretical models and see the dependence of HVS
velocity distributions on different ejection mechanisms, we first remove the
velocity deceleration (caused by the Galactic bulge, the disk, and the halo; see
below) from their observed values with correction for the proper motion and then obtain their velocities at infinity
by assuming that they move from the GC to the infinity only in the gravitational potential of the central
MBH. We denote these HVS velocities at infinity by $v^{\infty}_{\rm ej}$, and
their CDF $P(\geq v^{\infty}_{\rm ej})$ (hereafter, $v$CDF) represents the
number fraction of HVSs which have velocities at infinity higher than
$v^{\infty}_{\rm ej}$.

\begin{figure*}
\centerline{\includegraphics[width=4.5in]{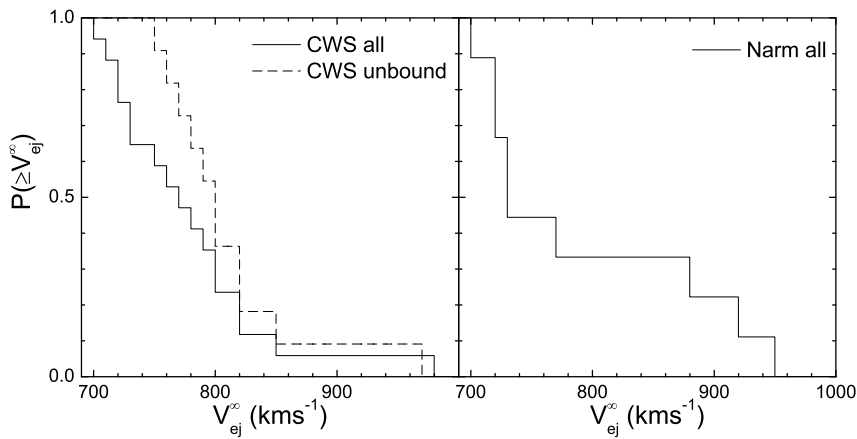}}
\caption{
Observational cumulative distribution of the ejection velocity at infinity
for the two populations of the detected HVSs ($v$CDF, $P(\geq v^{\infty}_{\rm
ej})$). Other legends are the same as in Figure~\ref{fig:f1}.
See Section~\ref{subsec:vd}.
}
\label{fig:f2}
\end{figure*}

Generally, the Galactic gravitational potential can be described by four
components, $\Phi=\Phi_{\rm BH}+\Phi_{\rm bulge}+\Phi_{\rm disk}+\Phi_{\rm
halo}$, i.e., the contribution from the central MBH, the Galactic bulge, the
disk, and the halo. We have $\Phi_{\rm BH}=-GM_{\bullet}/r$, where the mass of
the central MBH in the GC is adopted to be $M_{\bullet}=4\times 10^6\msun$
\citep[e.g.,][]{Ghezetal08,Gillessenetal09}. In this paper, we adopt the model
for the last three components from \citet{Xue08}, i.e., 
\be 
\Phi_{\rm bulge} = -\frac{GM_{\rm bulge}}{r+r_{\rm bulge}}, 
\ee 
with $M_{\rm bulge}=1.5\times10^{10}\msun$ and the core radius $r_{\rm
bulge}=0.6\kpc$,
\be 
\Phi_{\rm disk}  = -\frac{GM_{\rm disk}(1-e^{-r/b})}{r}, \label{eq:fdisc}
\ee 
with $M_{\rm disk}=5\times10^{10}\msun$ and the scale length $b=4\kpc$, and
\be 
\Phi_{\rm halo} = -\frac{4\pi G\rho_{\rm s} r^3_{\rm vir}}{c^3r} 
\ln(1+\frac{cr}{r_{\rm vir}}),
\ee
with $\rho_{\rm s}=\frac{\rho_{\rm c}\Omega_{\rm m}\Delta_{\rm vir}}{3}
\frac{c^3}{\ln(1+c)-c/(1+c)}$, where $\rho_{\rm c}$ is the cosmic critical
density, $\Delta_{\rm vir}=200$, $\Omega_{\rm m}$ is the cosmic fraction of
matter, the virial radius $r_{\rm vir}=267\kpc$, and the concentration $c=12$.
To see the velocity deceleration of a star caused by the bulge, disk and halo
potentials above, the minimum velocity required for the star starting from the
potential center (in the absence of the MBH) is $\sim645 \kms$ 
(or $670 \kms$, $700 \kms$) so that they can move to a distance of
$50\kpc$ (or $100\kpc$, $200\kpc$) further away.  With the Galactic potential
model above, we correct the velocity deceleration and obtain the $v$CDFs for
both HVS populations, which are shown in Figure~\ref{fig:f2}. 

In the estimation of $v^{\infty}_{\rm ej}$ above, we adopted the spherical
approximation of the Galactic disk potential given by Equation (2), since all
the detected HVSs are moving at Galactocentric distances much larger than the
disk scale length $b$. If using the Miyamoto$-$Nagai potential \citep{MN75} to
describe the flattened potential of the Galactic disk, i.e.,
$\Phi(R,z)=-GM_{\rm disk}/(R^2+(a_{\rm disk}+(z^2+b^2_{\rm
disk})^{1/2})^2)^{1/2}$ and $M_{\rm disk}=5\times 10^9\msun$, the scale length
and height of the disk are $a_{\rm disk}=5$\kpc, $b_{\rm disk}=0.4$\kpc,
respectively, we find that the difference in the estimated $v^{\infty}_{\rm
ej}$ due to the disk flattening is negligible ($\sim10 \kms$).

Adopting a different Galactic potential model may result in a different $v$CDF.
For example, choosing the Galactic potential as that given by Equation (8) in
\citet{Kenyon08}, the resulted $v^{\infty}_{\rm ej}$ ranges from $850 \kms$ to
$1200 \kms$, 
which is substantially higher than
that shown in Figure~\ref{fig:f2} (from $690 \kms$ to $980 \kms$). However,
the slope of the $v$CDF in this
velocity range is only slightly flatter than that estimated from
the potential given by (\citealt{Xue08}; also see discussions in
\citealt{Sesana07}).
Therefore, our conclusions made in
Sections~\ref{sec:Result} and \ref{sec:BBH} on comparisons of the simulated
$v$CDF (mainly the shape) with the estimated $v$CDF here are not sensitive to
the choice of the Galactic potential (see further discussions in
Section~\ref{subsubsec:velTBK}).

Note here that the Galactocentric distances of the detected HVSs range from
$25\kpc$ to $130\kpc$, which correspond to their travel time from the GC
ranging from a few tens to $260$~Myr.

\section{Tidal breakup of binary stars bound to the MBH}\label{sec:TBK}

The progenitors of the detected HVSs are probably originated from disk-like
stellar structures in the GC (e.g., the CWS disk) as demonstrated by
\citet{Luetal10}. These progenitors are likely to be bound to the central MBH.
In order to investigate whether the statistical distribution of the detected
HVSs shown in Section~\ref{sec:obs} can be reproduced by the TBK mechanism,
here we first explore the detailed physics of the tidal breakup of binary stars
that initially bound to the central MBH as it is still elusive in the
literature. Consider that the binary star is moving on an eccentric orbit bound
to the MBH and injecting/migrating into the vicinity of the MBH.  For convenience, we
denote the injecting/migrating stellar binary by the ``stellar'' binary with semimajor axis
$a_{\rm b}$, eccentricity $e_{\rm b}$, and mass $m=m_1+m_2$ (where $m_1$ and
$m_2$ are the masses of its two components, respectively). The center of mass
of the stellar binary and the MBH also compose a binary system, which we denote
by the ``outer'' binary with semimajor axis $a_{\rm out}$ and eccentricity
$e_{\rm out}$.

\subsection{Three-body experiments}\label{subsec:threebody}

We perform three-body experiments of dynamical interactions between stellar 
binaries and the central MBH.  We use the code DORPI5 based on the explicit
fifth(fourth)-order Runge$-$Kutta method \citep{DP80,Hairer93} to integrate the
set of the equations that control the three-body interactions between the stellar
binary and the MBH. We set the tolerance of the fractional energy error to be
$10^{-9}$ for a single orbit integration.
We have also checked that statistically our results are not affected by 
setting a smaller minimum tolerance of the energy error. 
The initial conditions set for the stellar and the outer
binaries include (1) the masses of the two components of the stellar binary
($m_1$ and $m_2$, $m_2\leq m_1$) and its mass ratio $q=m_2/m_1$; (2) the
initial semimajor axis and eccentricity of the stellar binary ($a_{\rm b,i}$,
$e_{\rm b,i}$); (3) the initial semimajor axis and eccentricity of the outer
binary ($a_{\rm out,i}$, $e_{\rm out,i}$);
(4) the orbital orientations of both the
stellar and outer binaries; and (5) the initial orbital phases of both the stellar and
outer binaries.  

For convenience, we also define a dimensionless penetration parameter $D$ to
describe how close a stellar binary can approach the MBH relative to its tidal
breakup radius $r_{\rm tb}$ (see a similar definition in \citealt{Hills88,
Bromley06}), i.e.,
\begin{equation}
D\equiv100\frac{r_{\rm p,i}}{r_{\rm tb}}=\frac{r_{\rm p,i}}{a_{\rm b,i}}
\left[\frac{3M_{\bullet}}{10^6(m_1+m_2)}\right]^{-1/3},
\end{equation}
where $r_{\rm p,i}=a_{\rm out,i}(1-e_{\rm out,i})$ is the initial pericenter
distance and $r_{\rm tb}=[3M_{\bullet}/(m_1+m_2)]^{1/3} a_{\rm b,i}$.
The penetration parameter
$D=100$ corresponds to the tidal breakup radius of a stellar binary, and
the range of $D$ of interest in this paper is $D \la 250$ where tidal breakup
of stellar binaries is possible.

\subsubsection{Initial settings}

For the numerical experiments in Sections~\ref{subsec:first} and
\ref{subsec:Multi}, we simply set the mass ratio of the stellar binary as $q=1$,
and $m_1=m_2=3\msun$; but for those in Section~\ref{sec:Result} we adopt more
realistic distributions of the mass ratio and the masses of the two components
of the stellar binaries according to current observations on binary stars.

We set the initial semimajor axis $a_{\rm b,i}$ of the stellar binary in the
range from $0.05\AU$ to $2\AU$.\footnote{The timescale for those stellar binaries to be
disrupted by their interactions with background stars is roughly $10^9$~yr 
\citep{YT03,Hopman09}. If these binaries are rotating around the MBH with semimajor 
axis of $\la 1$~pc as adopted in this paper, the total time for $\sim1000$ orbits is 
roughly $\la 10^7$~yr. Therefore, the evolution of stellar binaries due to their 
dynamical interactions with background stars can be safely ignored. }
Perturbations on a binary star with $a_{\rm
b,i}<0.05\AU$ from the tidal field of the MBH easily lead to the merger of its
two components; while tidal breakup of binary stars with $a_{\rm b,i}>2\AU$
usually leads to ejections of stars with velocities substantially smaller than
the hypervelocities interested in this paper.  For demonstration only, we
simply choose $a_{\rm b,i}=0.1\AU$ and $e_{\rm b,i}=0$ (or 0.1, 0.3, and 0.6 alternatively) in
the three-body experiments presented in Sections~\ref{subsec:first} and
\ref{subsec:Multi}; but in Section~\ref{sec:Result}, we adopt a distribution of
$a_{\rm b,i}$ based on the constraint obtained from current observations and
$e_{\rm b,i}=0$.

We set the semimajor axis of the outer binary as $a_{\rm out,i}\sim$0.04$-$0.5$\pc$ \citep{LuJ09,Gillessenetal09}, as the stellar binaries are
possibly originated from the CWS disk within half a parsec from the central
MBH. The pericenter distance of the binary is set to be close to the tidal
radius of the stellar binary. In Sections~\ref{subsec:first} and
\ref{subsec:Multi}, we set the penetration parameter $D$ typically to be $\sim
$20$-$250 and choose $a_{\rm out,i}=0.2\pc$ for demonstration only.

The orientation of the orbit of the stellar binary, relative to that of the outer
binary, $\phi\in[0,\pi]$, is assumed to be uniformly distributed in
$\cos\phi$, if not specified. 

The initial orbital phases of both the stellar and outer binaries are randomly
chosen in all the following calculations.

\subsubsection{Approximations}\label{subsubsec:approx}

As the stellar binaries are initially set on orbits weakly bound to the
central MBH, binaries with large penetration parameters (e.g., $D\ga150$) may
revolve around the central MBH for many (e.g., 10$-$100 or even $1000$) orbits.
In our calculations below, the period of the stellar binary is usually much
smaller than that of the outer binary.  Therefore, most of the calculation time
may be spent on integrating the stellar binary orbit when it is faraway from the
MBH.  However, when the binary star is faraway from the MBH, the tidal torque
from the MBH on the binary star is rather weak, so we can approximate the
motion of the stellar binary around the MBH into two independent two-body
problems: one is for the stellar binary, and the other is for the outer binary on
an elliptical orbit, both of which can be done analytically. We adopt the above
two-body approximation when the tidal force from the MBH on the stellar binary is
less than $10^{-6}$ of the gravitational force between its two components (we have
checked the two-body approximation by setting a lower threshold, e.g., $10^{-7}$ 
or $10^{-8}$, and found no significant difference in our results).
With this approximation, our calculation time is sped up substantially when
$a_{\rm out}$ is large ($\ga 0.2\pc$) and its accuracy can still be maintained.

A star may be tidally disrupted if its close passage to the MBH is $\la
r^*_{\rm tid}\equiv R_*(M_{\bullet}/m_*)^{1/3}$, where $m_*$ and $R_*$ are the mass
and radius of the star, respectively. The radius of a star can be roughly
given by $R_*\propto R_{\odot}(m_*/M_{\odot})$ for $m_*<M_{\odot}$ and
$R_*\propto R_{\odot}(m_*/ M_{\odot})^{0.75}$ otherwise, where $R_{\odot}$ is
the solar radius \citep[][and references therein]{Torres09}. Due to the tidal
disruption, part of the disrupted remnants may be swallowed by the central MBH
and part may be ejected out. In our calculations, we terminate the evolution of
the corresponding system once any component of the stellar binary approaches
within a radius $r^*_{\rm tid}$ from the MBH.

The tidal torque from the MBH may change the semimajor axis and the
eccentricity of the stellar binary during its close passage to the MBH (see
further discussions in Section~\ref{subsubsec:bound}). In the three-body
experiment, we assume that the two components of the binary star merge into a
single star once the distance between the two components becomes smaller than
the sum of their radii, and the evolution of the corresponding system is
terminated once a merger event occurs.

The consequences on the dynamical interactions of stellar binaries with the
MBH can be different depending on whether the stellar binaries are initially
on bound or unbound orbits. If the binary star is initially bound to the MBH,
it may revolve around the central MBH for many orbits before its disruption or
merger; while the binary initially unbound to the MBH passes by the MBH only
once even if it is not broken up during its close passage (see further
discussions in Sections~\ref{subsec:first} and \ref{subsec:Multi}).  Multiple
times of interactions of bound binary stars with a central MBH have been
discussed by \citet{Antonini09}, in which they focus on the post-Newtonian
effects on stellar orbits in the gravitational field of the MBH and mergers of
the two components of the binary.  In Section~\ref{subsec:first}, we analyze
the consequence of the {\it first} close encounter of the stellar binary with the
central MBH, and the changes in the orbital elements of the stellar binary (if
the binary star survives). Then we consider the cumulative effects of multiple
encounters in Section~\ref{subsec:Multi}. We focus on discussing how the
spatial and velocity distributions of HVSs are connected with their
originations, ejecting mechanisms, and dynamical environments in the GC.

In Section~\ref{subsec:first}, $10^4$ three-body experiments are performed for
each of the four set of initial conditions, i.e., $e_{\rm b,i}=0$, $0.1$, $0.3$,
and $0.6$, respectively. If a stellar binary is not tidally broken up during its
first close passage to the MBH, the three-body experiment ends up when the 
stellar binary reaches the apoapsis
after its first close passage to the MBH.  In Section~\ref{subsec:Multi},
$10^4$ three-body experiments are performed for each set of initial conditions
and those experiments that do not lead to disruption/merger/separation within
$500$ revolutions are excluded.

\subsection{Consequences of the first close passage of a stellar binary to
the MBH}\label{subsec:first}

Considering that binary stars are injected toward the MBH on weakly bound
orbits with eccentricity close to 1 and $D\la 250$, the changes in their orbits,
as consequences of the tidal effect from the MBH, 
after their first close passages, can be characterized 
by the following five categories.

\begin{enumerate}

\item The two components of the stellar binary remain bound to each other, but its
orbital parameters ($a_{\rm b}$, $e_{\rm b}$) are changed by the tidal
torque from the MBH during the close encounter. The outer binary is still
bound and the stellar binary will encounter with the MBH again in its next close
passage. The probability of our three-body experiments below resulted in this category is denoted by $P_{\rm bound}$. 

\item The binary star is tidally broken up, with one component of the binary
being ejected out as a high-velocity star, and the other one being trapped onto an
orbit that is tightly bound to the MBH compared to its parent stellar binary.
The probability resulted in this category is denoted by $P_{\rm ej}$.

\item The binary is tidally broken up into two single stars, and the two stars
are not bound to each other but both bound to the central MBH.
The probability resulted in this category is denoted by $P_{\rm sp}$.

\item The two components of the binary merge into one single star.
The probability resulted in this category is denoted by $P_{\rm mrg}$.

\item One (or both) component(s) of the binary is (are) tidally disrupted
and partly swallowed by the central MBH. 
The probability resulted in this category is denoted by 
$P_{\rm swallow}(=1-P_{\rm bound}-P_{\rm ej}-P_{\rm sp}-P_{\rm mrg})$.

\end{enumerate}

\subsubsection{Consequence 1 }\label{subsubsec:bound}

If a stellar binary can still maintain its integrity after its close
passage to the central MBH, the eccentricity and the semi-major axis of the
binary are generally changed due to the tidal torque from the MBH.

\begin{itemize}

\item {\it Changes in eccentricities}. Figure~\ref{fig:f3} shows the simulation
results of the changes in eccentricities of the stellar binaries after their
first close passages to the MBH. When the penetration parameter is small (e.g.,
$D\la 150$), the eccentricity of the stellar binaries may be excited (or
de-excited) to various values in the range from 0 to 1. The large range of the
excited (or de-excited) eccentricities is mainly due to various phases of the
stellar binaries at the close passages to the MBH and various orientations of the
stellar binaries relative to the orbit of the outer binary.  For encounters with
$D \ga 200$, excitation (or de-excitation) on eccentricities is relatively
small, especially for those binaries with $e_{\rm b,i}=0$.  Multiple subsequent
encounters of the stellar binary with the MBH should be common at $D \ga 150$,
where both $P_{\rm ej}$ and $P_{\rm mrg}$ are small and $P_{\rm bound}$ is large
(see Figure~\ref{fig:f5}). In these cases, the binary eccentricity (and
semimajor axis) may then be further changed due to the cumulative effect of
multiple subsequent encounters (see Section~\ref{subsec:Multi}).

\begin{figure*}
\centerline{\includegraphics[width=6.0in]{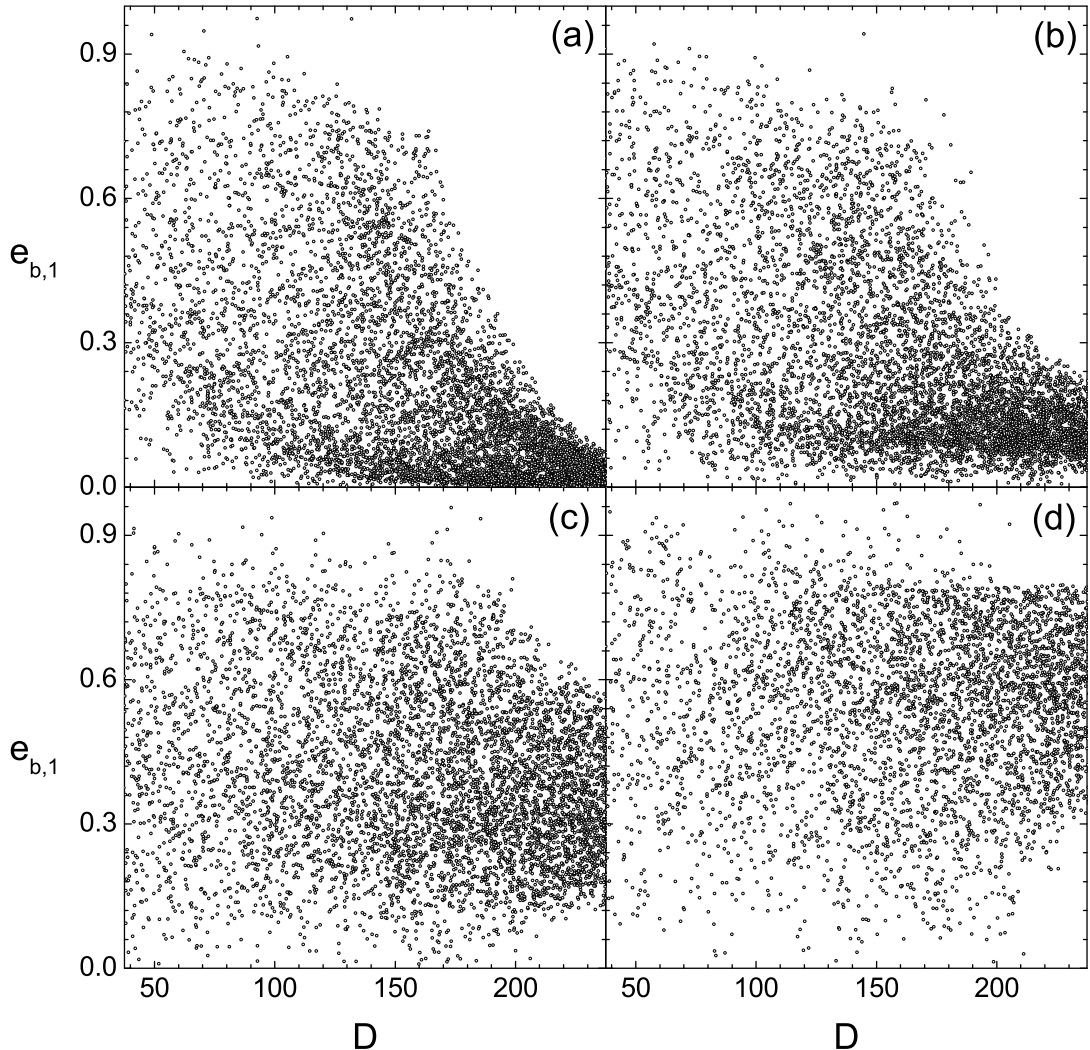}}
\caption{Eccentricity of the binary stars after their first close
encounters with the central MBH ($e_{\rm b,1}$) vs. the penetration
parameter $D$, if the binary star survives.  The semimajor    
axes and the masses of the binaries are all set to be $a_{\rm b,i}=0.1\AU$ and     
$m_1=m_2=3M_{\odot}$, respectively. The semimajor axes of the outer binaries       
are all set to be $a_{\rm out,i}=0.2\pc$.  Panels (a)-(d) show the $e_{\rm b,1}$ for those stellar binaries with 
initial eccentricities of $e_{\rm b,i} =0$, 0.1, 0.3, and 0.6, respectively. As
seen from this figure, the eccentricity of a binary may change to various values
between 0 and 1 after the first close encounter if $D<150$. The total numbers of 
survived binary stars in the $10^4$ three-body experiments here are $7256$, 
$7160$, $6709$, and $4782$ in panels (a), (b), (c), and (d), respectively. }
\label{fig:f3}
\end{figure*}

\item {\it Changes in semimajor axes}. As shown in Figure~\ref{fig:f4}, the semimajor
axis of a binary star can be excited to a larger value but can also be
de-excited to a smaller value. The changes of the semimajor axes depend on the
orbital phases of the stellar binary during its close encounter. If the relative
positional vector of the two components of the stellar binary is aligned with the
radial vector from the central MBH, the stellar binary is more likely to be
stretched and its semimajor axis is changed to a larger value after the
encounter; while the binary can also be shrunk and its semimajor axis may be
changed to a smaller value if the relative positional vector of its two
components is close to perpendicular to the radial vector. The time for the
binary passing by the periapsis is comparable to, if not substantially smaller
than, the orbital period of the stellar binary when the closest approach is
roughly the tidal radius of the binary. Therefore, whether the binary is
stretched or shrunk is determined by the orbital phase and orientation of the
stellar binary during its close encounter with the central MBH. As the stellar
binary is stretched in most of its orbital phases, the semimajor axis of the
binary is more likely to be excited rather than de-excited after its first
close passage to the MBH (see Figure~\ref{fig:f4}).  We also find that the
relative change of the semimajor axis of the stellar binary is independent of the
initial semimajor axis of the binary, which is verified by our calculations
with adopting different $a_{\rm b,i}$. 

\begin{figure*}
\centerline{\includegraphics[width=6.0in]{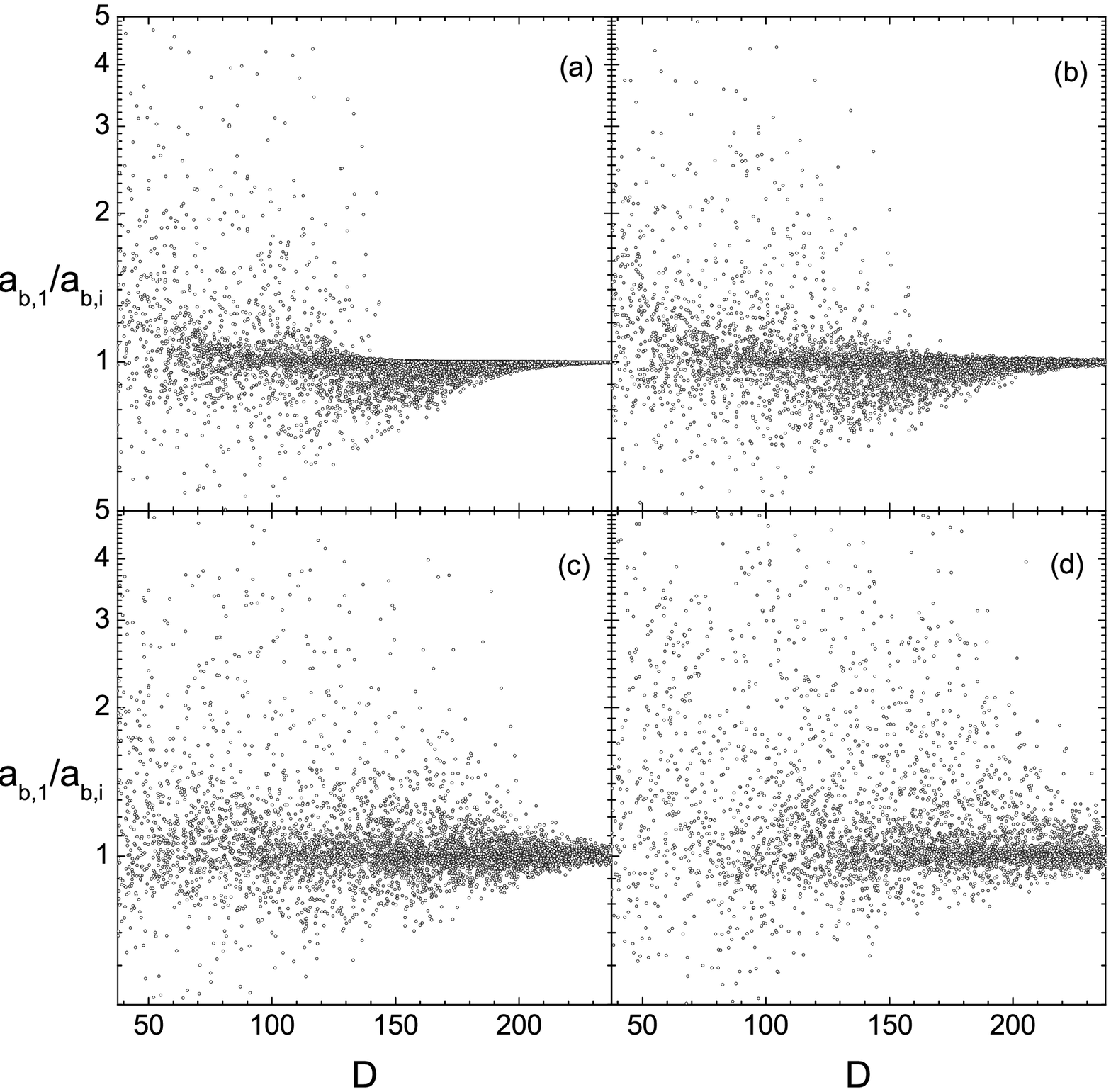}}
\caption{Relative changes in the semimajor axes of the binary stars after their
first close passages to the central MBH as a function of the penetration
parameter $D$. The initial conditions for the stellar binaries are the same
as that in Figure~\ref{fig:f3}. The initial eccentricities of the stellar
binaries are $e_{\rm b,i}=0$ (panel (a)), 0.1 (panel (b)), 0.3 (panel 
(c)), and 0.6 (panel (d)), respectively. Generally, the change in 
$a_{\rm b}$ is large if $D$ is small. The total number of survived binary stars shown
in each panel is the same as that in Figure~\ref{fig:f3}.
} 
\label{fig:f4} 
\end{figure*}

In addition, we note here that the orbit of the outer binary star 
almost remains the same after the first close encounter if the stellar
binary can maintain its integrity.  Our calculations above show that the           
semimajor axis of outer binaries may change by $\delta a_{\rm out}\la              
0.02a_{\rm out}$ and the distance to the pericenter of the outer binary may        
only change by $\delta r_{\rm p}\la 0.001 r_{\rm p}$, which suggests that the      
location of the subsequent close encounters is almost the same as that of the      
first encounter if no other perturbation takes place on the orbit of the binary    
star.                                                                              
                                                                                   
\end{itemize}                                                                      

\subsubsection{Consequences 2 and 4: ejection and merging probabilities}
\label{subsubsec:ejectionmerger}

\begin{figure}
\epsscale{1.0}
\plotone{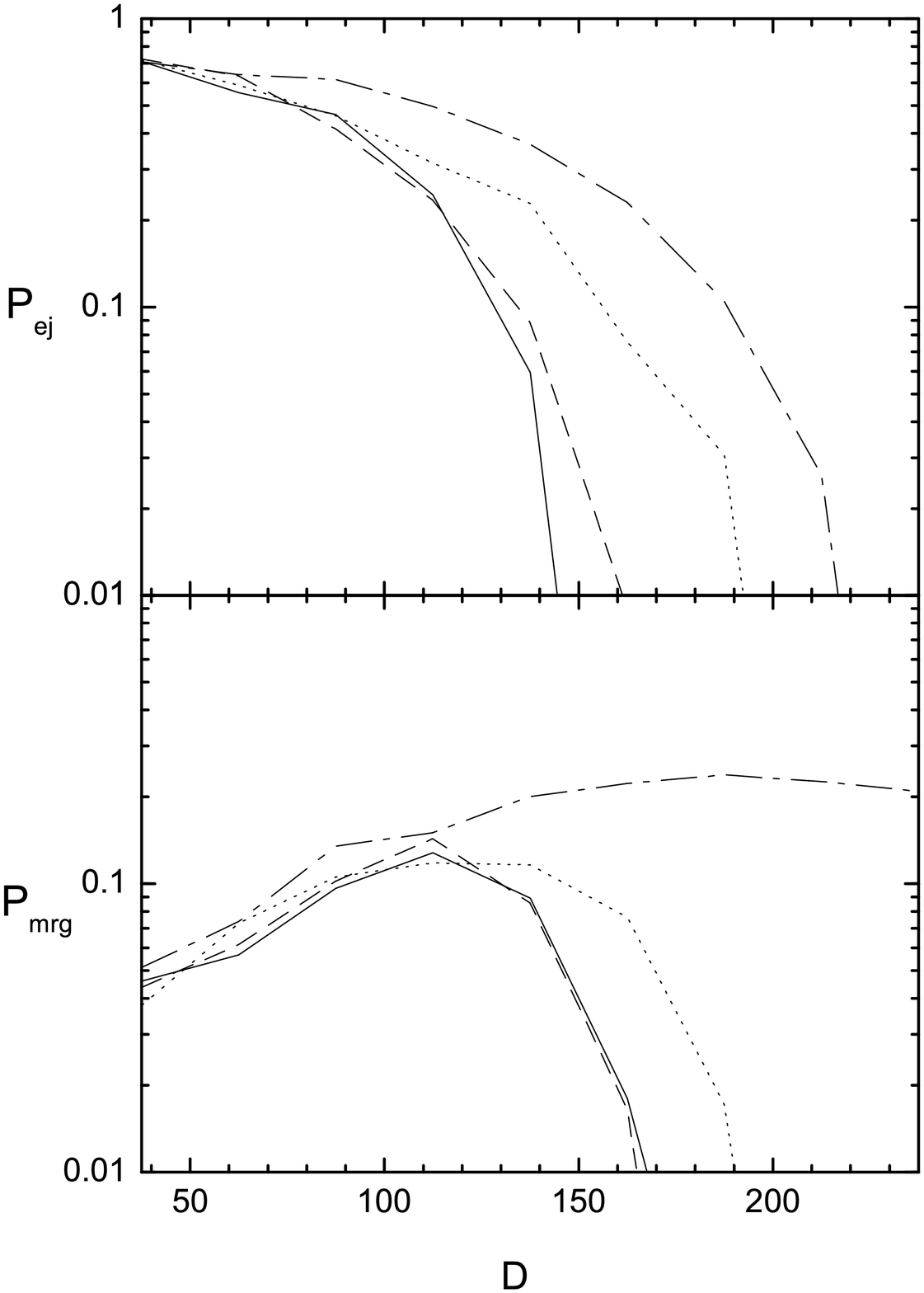}
\caption{
Ejection and merging probabilities ($P_{\rm ej}$ and $P_{\rm mrg}$) of the
stellar binaries after their first close passages to the MBH as a function of
the penetration parameter $D$ (top panel and bottom panel, respectively). 
The initial conditions set for the stellar binaries are the same as those in Figure~\ref{fig:f3}.
The solid, dashed, dotted, and dot-dashed curves represent the results for those
stellar binaries with initial eccentricities $e_{\rm b,i}=0$, $0.1$, $0.3$, and $0.6$,
respectively. }
\label{fig:f5}
\end{figure}

Figure~\ref{fig:f5} shows our calculation results on the ejection probability
$P_{\rm ej}$ and the merging probability $P_{\rm mrg}$. As seen from Figure~\ref{fig:f5},
$P_{\rm ej}$ increases with decreasing penetration parameters $D$ because
the tidal force from the MBH increases with decreasing $D$. And $P_{\rm ej}$ also increases with increasing
initial eccentricities of the stellar binaries. The reason is that 
the distance between the two components of a binary for given $D$ and $a_{\rm b}$, and hence the tidal
torque on them, is larger if the binary has a larger eccentricity, and therefore the binary is easier to be broken up. For the same reasons, $P_{\rm mrg}$ shown in
Figure~\ref{fig:f5} decreases with increasing $D$ at the large-$D$ end, but
decreases with decreasing $D$ at the low-$D$ end due to the significant
increase of $P_{\rm ej}$. A stellar binary with a larger initial eccentricity
is also easier to be excited to an eccentricity high enough for its two
components to merge and thus $P_{\rm mrg}$ increases with increasing $e_{\rm
b,i}$ at large-$D$ end.

\begin{figure*}
\centerline{\includegraphics[width=6.0in]{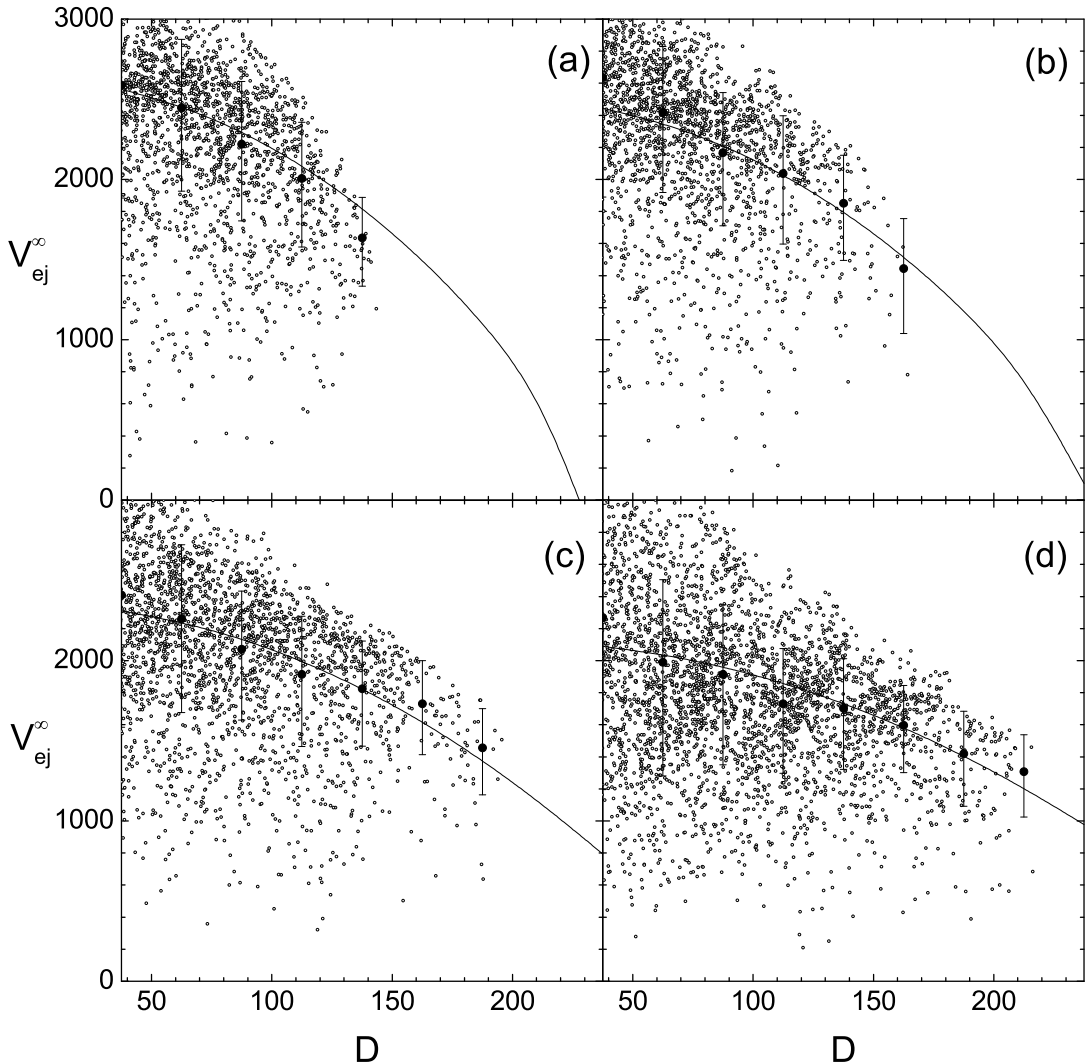}}
\caption{
Ejection velocity at infinity ($v^{\infty}_{\rm ej}$) for the ejected
components of those stellar binaries that are tidally broken up after their
first close passages to the central MBH.  The initial conditions for the
stellar binaries are the same as that in Figure~\ref{fig:f3}. Panels (a),
(b), (c), and (d) show the results for those stellar binaries with initial
eccentricities $e_{\rm b,i}=0$, 0.1, 0.3, and 0.6, respectively. The filled
circles are the square roots of the rms values of $(v^{\infty}_{\rm ej})^2$ in different bins of
penetration parameters $D$, and the error bars indicate the square roots of the standard deviations of $(v^{\infty}_{\rm ej})^2$.
The solid curves show the best fits to the rms values. The total
numbers of the experiments shown are $2261$, $2339$, $2687$, and $3543$ in
panels (a), (b), (c), and (d), respectively.
}
\label{fig:f6}
\end{figure*}

Figure~\ref{fig:f6} shows the ejection velocity at infinity ($v^{\infty}_{\rm
ej}$) of the ejected component as a function of the penetration parameter ($D$)
of stellar binaries with different initial eccentricities ($e_{\rm b,i}=0$,
$0.1$, $0.3$, and $0.6$, respectively). We find that the rms values of the
ejection velocities can be well fitted by the following formula:
\begin{eqnarray}
\langle (v^{\infty}_{\rm ej})^2\rangle^{1/2}&=&\frac{v_{\rm ej,0}^{\infty}}{\sqrt{1+e_{\rm b,i}}}
\left[1-\left(\frac{D_{\rm eff}}{260}\right)^2\right] \left(\frac{a_{\rm
b,i}}{0.1\AU}\right)^{-1/2} \times \nonumber \\ &
&\left(\frac{m_1+m_2}{6\msun}\right)^{1/3} \left(\frac{M_{\bullet}}{4\times
10^6\msun}\right)^{1/6},     
\end{eqnarray}
for $20<D<260$, where $D_{\rm eff}=D/\sqrt{1+e_{\rm b,i}}$ is the effective
penetration parameter for a stellar binary with non-zero eccentricity, and
$v^{\infty}_{\rm ej,0}=2560 \kms$ is the normalization of the ejection velocity
when $e_{\rm b,i}=0$ and $a_{\rm b,i}=0.1\AU$. For stellar binaries with zero
initial eccentricities and any $D$ in the range of 20$-$150, this fitting formula gives a value
similar to that given by Equation (1) in \citet{Bromley06}.

If $a_{\rm out,i}$ is set to a smaller value, the ejection velocity may be
changed by $v^{\infty}_{\rm ej}\rightarrow \sqrt{(v^{\infty}_{\rm
ej})^2-GM_{\bullet}/a_{\rm out,i}}$ because the ejected star has to first
overcome its initial bounding energy. If $a_{\rm out,i}=0.01\pc$, for example,
the ejection probability $P_{\rm ej}$ is significantly smaller than that shown
in Figure~\ref{fig:f5} and the ejection velocity $v^{\infty}_{\rm ej}$ is also
substantially smaller because a significant fraction of the stellar binaries
end up in consequence 3 instead of consequence 2.  

\subsection{Multiple encounters}\label{subsec:Multi}

\begin{figure} \epsscale{1.0} \plotone{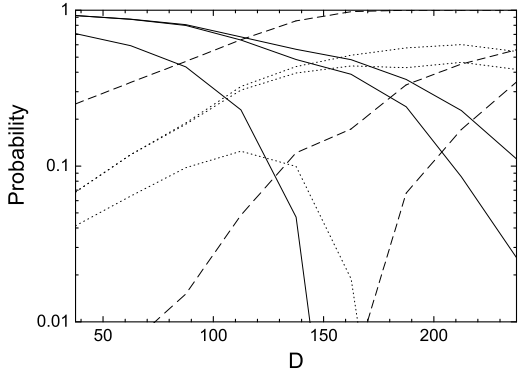} 
\caption{Probabilities of different consequences of binary stars after
$N_{\rm orb}$ times close passages to the central MBH as a function of the
initial penetration parameter $D$. The solid, the dotted, and the dashed curves
represent $P_{\rm ej}$, $P_{\rm mrg}$, and $P_{\rm bound}$, respectively (see
Section~\ref{subsec:first}). 
For each type of the curves, we show the results for $N_{\rm orb}=1$, 20, and 500, respectively (solid and dotted curves: from
bottom to top; dashed curves: from top to bottom).} \label{fig:f7} \end{figure}

\begin{figure} \epsscale{1.0}
\plotone{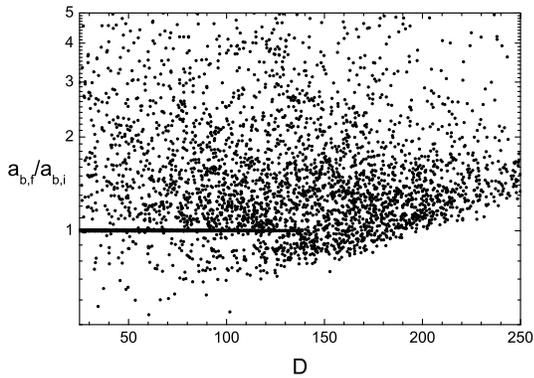} \caption{Ratio of the semimajor axis of the stellar
binary to its initial value ($a_{\rm b,f}/a_{\rm b,i}$) vs. the penetration
parameter $D$, where $a_{\rm b,f}$ is the semimajor axis of the stellar binary
when it is at the apocenter of the last outer binary orbit just before its
breakup. For those stellar binaries with $D>150$, most of them have $a_{\rm
b,f}$ substantially larger than $a_{\rm b,i}$. In the figure, the total number
of the stellar binaries that are broken up is 6007, of which 2116 are
broken up after their first close passages to the MBH and thus have
$a_{\rm b,f}/a_{\rm b,i}=1$.
 } \label{fig:f8} \end{figure}

\begin{figure}
\epsscale{1.0} 
\plotone{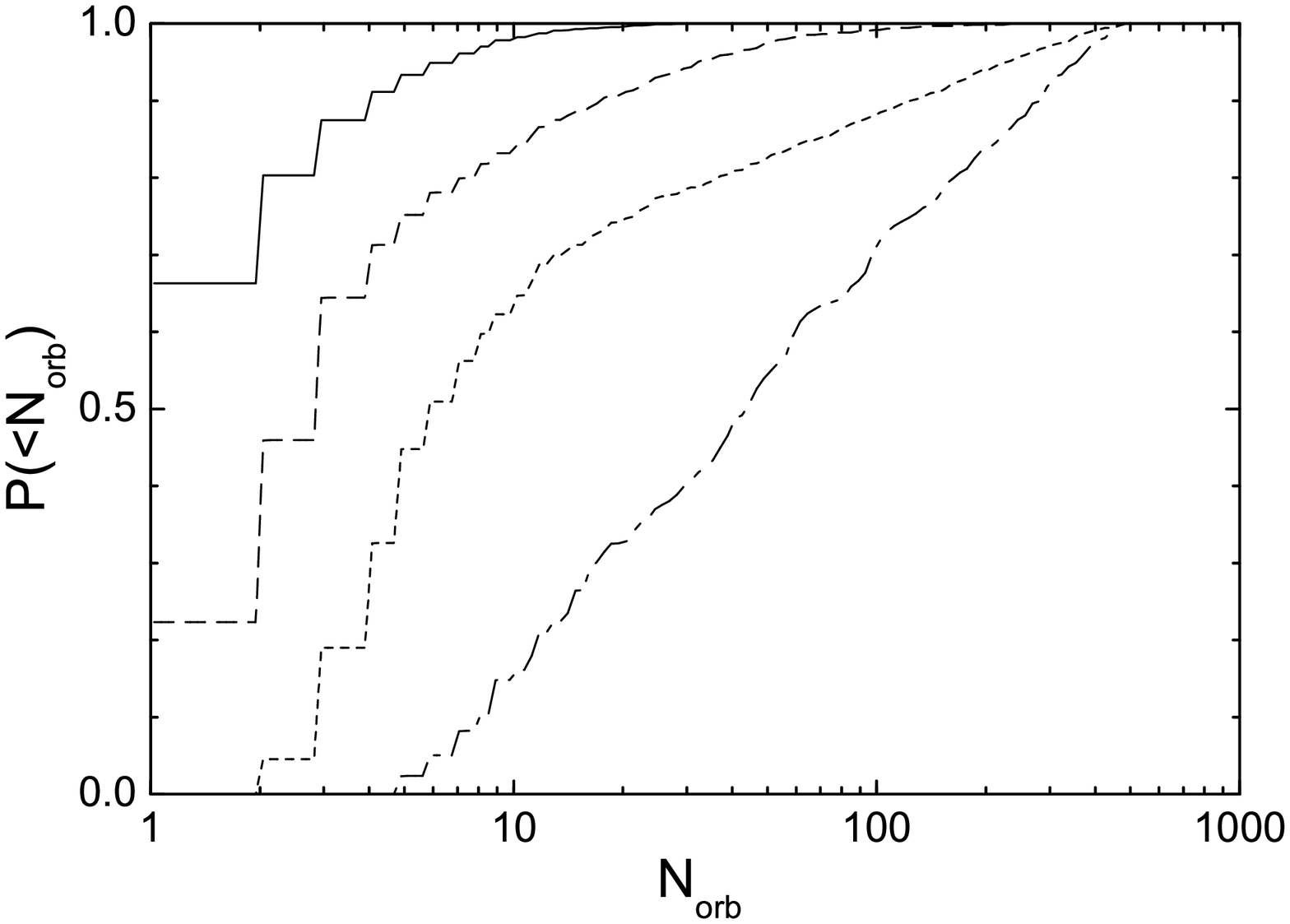} \caption{Cumulative probability distributions of the
number of close passages that the stellar binaries have experienced before
their breakup with ejecting one of their two components. Only those stellar
binaries that are broken up within 500 revolutions around the MBH are 
taken into account.  
The solid, dashed, dotted, and dot-dashed lines represent the distributions
with different ranges of penetration parameters $25<D<100$, $100<D<150$,
$150<D<200$, and $200<D<250$, respectively.} \label{fig:f9} \end{figure}

\subsubsection{The ejection/merging probabilities and the number of
revolutions}

The tidal effect from the MBH may be not large enough to break up a stellar
binary with a large $D$ (e.g., $D\ga 150$) during the first close passage of
the binary. If the binary star is initially bound to the MBH, multiple times of
close passages of the binary to the MBH should be common, and thus the
probabilities of ejection and stellar coalescence are substantially enhanced,
especially at large $D$, as demonstrated in Figure~\ref{fig:f7}, which can be
understood as follows. 

\begin{itemize}

\item {\it The increase of the number of close passages to the MBH}. Assuming that
the ejection probability after each close passage of the binary is roughly the
same as denoted by $P_{\rm ej,0}$ and the probability of the binary maintaining
its integrity is roughly $P_{\rm bound,0}$, then after $N_{\rm orb}$ revolutions
the cumulative ejection probability is given by
\begin{equation}
P_{\rm ej}\simeq \sum_{i=1}^{N_{\rm orb}} P_{\rm ej,0}P_{\rm bound,0}^{i-1}
=P_{\rm ej,0} \frac{1-P_{\rm bound,0}^{N_{\rm orb}}}{1-P_{\rm bound,0}}.
\end{equation}
If $N_{\rm orb}\rightarrow\infty$, then $P_{\rm ej}=P_{\rm ej,0}/(1-P_{\rm
bound,0})>P_{\rm ej,0}$. When $D$ is large ($\ga 150$), $P_{\rm ej}$ can
increase substantially with increasing $N_{\rm orb}$ as $P_{\rm bound,0}$ is
close to 1.

\item {\it The increase of the semimajor axis of the stellar binary}. The semimajor
axis of the stellar binary may be excited to be larger than its initial value
after the close passages (see Figure~\ref{fig:f6}), and correspondingly the
penetration parameter $D\propto r_{\rm p}/a_{\rm b}$ decreases as $r_{\rm p}$
is almost not changed. A smaller $D$ results in a larger $P_{\rm ej}$. As
demonstrated in Figure~\ref{fig:f8}, the semimajor axes of many stellar binaries
before their breakup and last close encounters with the MBH are substantially
larger than their initial values. As the tidal force is approximately a
monotonically increasing function of $a_{\rm b}$, the semimajor axes of many
binaries tend to rapidly (exponentially) increase after many subsequent close
encounters with the MBH, while the semimajor axes of some other binaries
decrease only slightly, which leads to the asymmetric distribution of $a_{\rm
b,f}/a_{\rm b,i}$, i.e., the relative change in $a_{\rm b}$ for many binaries
with $a_{\rm b}$ excited is substantially larger than that for a relatively small
number of binaries with $a_{\rm b}$ de-excited, as shown in
Figure~\ref{fig:f8}.

\item {\it The increase of the eccentricity of the stellar binary}. Binary stars with
large penetration parameters ($D\ga150$) may be hard to break up initially.
However, their eccentricities are relatively easy to be excited to large values
due to cumulative tidal effects from the central MBH during their multiple
encounters with the MBH. The increase in the eccentricity thus leads to a
larger ejection or merging probability as mentioned in
Section~\ref{subsec:first}. We note here that the Kozai mechanism may cause
periodical oscillations of the internal eccentricity of stellar binaries during
their multiple encounters with the MBH, which may be important in enhancing the
merger of the two components of those binaries. A detailed study of this resonance
can be found in \citet{Antonini09}.

\end{itemize}

For those stellar binaries injected into the vicinity of the MBH with $D\sim$200$-$250, $\sim90\%$ of them are disrupted or merged within $500$ revolutions.
For those stellar binaries that are broken up within $500$ revolutions with
one component being ejected out, Figure~\ref{fig:f9} shows the cumulative
probability distribution of the number of revolutions around the MBH ($N_{\rm
orb}$) that they experienced. The majority of the stars are broken up within
10 revolutions when $D$ is small, while most of them can be broken up only
after many revolutions when $D$ is large. If $250>D>200$, for example, $N_{\rm
orb}$ ranges from 4 to 500 (see Figure~\ref{fig:f9}), and about 40\% of those
stellar binaries are broken up after 100 encounters. We should note that the 
ejection velocities of the resulted HVSs at large $D$ are substantially smaller
than those of the stellar binaries broken up at small $D$.

\subsubsection{The effects of relative orientations}

The tidal effects on a stellar binary during its close passages to the central
MBH are also affected by the initial orientation of the stellar binary orbit with
respect to the outer binary orbit. Given the penetration parameter $D$, the
prograde rotating stellar binary is easier to be tidally broken up and can
receive a larger kick in velocity during the close passages than a retrograde
rotating stellar binary. Compared to prograde rotating stellar binaries, retrograde
rotating binaries can revolve around the central MBH for many more orbits
before they are broken up. And the ejection velocities of the stars from the
interaction of retrograde binaries with the MBH are also much smaller than that
for the cases of prograde binaries. In Table~\ref{tab:tab1}, we list some
numerical results to show the effect of different relative orientations of the
stellar binary with respect to that of the outer binary. (For demonstration, only
the results from those experiments with the stellar binaries initially having
$a_{\rm b,i}=0.1\AU$, $e_{\rm b,i}=0$, and $D=100$ are shown in
Table~\ref{tab:tab1}.)

\begin{deluxetable}{lcccccc}
\tablecaption{Effects of the relative orientation of the stellar binary orbit
to the outer binary orbit}
\tablehead{\colhead{Orientation} & \colhead{$P_{\rm ej}$} &
\colhead{$P_{\rm mrg}$} & \colhead{$P_{\rm sp}$} 
& \colhead{$\sqrt{\langle {v^{\infty}_{\rm ej}}^2\rangle}$\tablenotemark{a}} & \colhead{$\frac{\sigma}{
\sqrt{\langle {v^{\infty}_{\rm ej}}^2\rangle}}$\tablenotemark{b}} & 
\colhead{$\left<N_{\rm orb}\right>$\tablenotemark{c}}}
\startdata
Prograde   & 0.88 & 0.12 & 0.000 & 2.39 & $^{+0.13}_{-0.15}$  & 1.1 \\
Retrograde & 0.66 & 0.29 & 0.038 & 1.17 & $^{+0.53}_{-\cdots}$    & 16  \\
Uniform    & 0.86 & 0.13 & 0.004 & 1.80 & $^{+0.27}_{-0.37}$  & 3.9 
\enddata
\label{table:t1}
\tablenotetext{a}{Mean ejection velocity at infinity in units of $10^3 \kms$.}
\tablenotetext{b}{Errors in the rms of the ejection velocities at infinity.}
\tablenotetext{c}{Mean number of close encounters experienced by the 
stellar binaries before their breakup and the ejection of one of its two
components (i.e., the consequence 2 in Section~\ref{subsec:first}).}
\tablecomments{
In the first column, ``prograde'' (or ``retrograde'') denotes that all the stellar
binaries are rotating in the same direction as (or the anti-direction of) the
outer binary, while ``uniform'' denotes that the orientations of the stellar binaries with
respect to that of the outer binary $\phi$ are uniformly distributed in
$\cos\phi$ for $\phi\in [0,\pi]$. This table shows the results from those experiments
with the stellar binaries initially having $a_{\rm b,i}=0.1\AU$, $e_{\rm b,i}=0$, 
and $D=100$. }
\label{tab:tab1}
\end{deluxetable}

\begin{deluxetable}{ccccc}
\tablecaption{Different models for the TBK mechanism}
\tablehead{\colhead{Model} & \colhead{$f(a_{b,i})$\tablenotemark{a}} & 
\colhead{$f(r_{\rm p})$} & \colhead{$a_{\rm out,i}(\pc)$}  & 
\colhead{$v^{\infty}_{\rm ini}(\kms)$\tablenotemark{b}} 
}
\startdata
UB-1  & \"{O}pik   & Constant    & $\cdots$   & $250$     \\
UB-2  & log-N      & Constant    & $\cdots$   & $250$     \\
LP-1  & \"{O}pik   & Constant    & 0.04$-$0.5 & $\cdots$  \\
LP-2  & log-N      & Constant    & 0.04$-$0.5 & $\cdots$  \\
RW-1  & \"{O}pik   & Random walk & 0.04$-$0.5 & $\cdots$  \\
RW-2  & log-N      & Random walk & 0.04$-$0.5 & $\cdots$  
\enddata
\tablenotetext{a}{The ``log-N'' denotes those models adopting a log-normal distribution
of $f(a_{\rm b,i})$ as that given by \citet{DM91}, and the ``\"{O}pik'' denotes those 
models adopting the \"{O}pik law for $f(a_{\rm b,i})$.}
\tablenotetext{b}{For the ``UB'' model, those stellar binaries initially have
velocities of $250 \kms$ at infinity.}
\label{tab:tab2}
\end{deluxetable}                                                  

\subsubsection{Deflection angles}\label{subsubsec:Deflection}

Here, we briefly summarize the results about the deflection angles between the
direction of an HVS moving away from the Galactic halo and the initial
injecting direction of its progenitorial binary. The ejecting direction of an
HVS is almost anti-parallel to the injecting direction of its progenitor and
the off-set $\delta \Theta$ is $\la 15\arcdeg$, as demonstrated by \citet[][see
Figure~1]{Luetal10}. In another words, HVSs can well memorize the injecting
directions of their progenitors. We also find here that the major component of
$\delta\Theta$ is in the orbital plane of the outer binary and the other
component of the off-set, which is perpendicular to the orbital plane of the
outer binary and is relatively insignificant $\sim1\arcdeg$$-$$2\arcdeg$. The main
reason for this result is that the stretching on the stellar binary by the MBH
tidal force is most effective when the orientation of the stellar binary is
similar to that of the outer binary orbit. Considering the gravitational
potential due to the stellar cusp surrounding the MBH (which currently seems not to be clearly determined; \citep{Doetal09,Bartko09b}), the orbit of a
stellar  binary revolving around the MBH may precess significantly if the
binary is broken up only after more than hundreds of revolutions. Due to this
precession, the ejecting direction of a resulted HVS may deviate significantly
from the injecting direction of its progenitor and the main component of the
deviation is in the orbital plane of the outer binary. If projecting these two
directions onto the plane that is perpendicular to the orbital plane of the
outer binary, however, the difference between the two projected directions is
still insignificant. This suggests that the spatial distribution of the
ejecting HVSs can still maintain on the plane that their progenitors originated
even if the precession due to the stellar cusp is significant. Therefore, the
spatial distribution of HVSs should reflect the geometrical structure of the
parent populations of their progenitorial binaries. If the progenitors of HVSs
are originated from a disk-like structure, particularly, the resulted HVSs
projecting to the infinity on the sky should locate close to a great circle
with the same orientation as the disk (see Figure~5 in \citealt{Luetal10}).  We
will demonstrate this further in Section~\ref{sec:Result}.

\section{Inclination and velocity distributions of HVSs from the TBK mechanism} \label{sec:Result}

\subsection{Initial settings}

To investigate the spatial and velocity distributions of HVSs, we now choose
more realistic distributions of the parameters for both the stellar and outer
binaries (i.e., initial conditions) involving in our numerical simulations as
follows. 

For the stellar binary, the initial conditions are set as follows under the
assumption that they are originated from stellar disk(s) like the CWS one, though
many important issues related to the disk origination of HVSs (such as warping and
relaxation of the disk(s) and many young stars observed in the GC but not in
the CWS disk; \citealt{LuJ09,Bartko09b}) need to be further investigated \citep{Luetal10}. (1) The
distribution of their semimajor axes $a_{\rm b,i}$ follows either the \"{O}pik
law, i.e., $P(a_{\rm b}) da_{\rm b} \propto da_{\rm b} /a_{\rm b}$
\citep[e.g.,][]{KF07}, or a log-normal distribution \citep{DM91}. (2) The mass
distribution of the primary stars $m_1$ follows the Miller$-$Scalo initial mass
function, i.e., $f(m_1) \propto m_1^{-\alpha}$ and $\alpha \sim 2.7$
\citep[e.g.,][]{Kroupa}.\footnote{We have tested that adopting a top-heavy
initial mass function as that in \citet{Bartko09b} does not significantly
affect the resulted spatial and velocity distributions.} For massive binary
stars, the distribution of the secondary star or the mass ratio $q=m_2/m_1$ can
be described by two populations: (a) a twin population, i.e., about 40\% binary
stars have $q\sim 1$ and (b) the rest binaries, which follow a distribution of
$f(q) \sim {\rm constant}$ \citep{KF07,Kiminki08,Kiminki09}.  (3) The initial
eccentricity of the stellar binary is assumed to be $0$, as adopted in some
previous works (e.g., \citet{Bromley06,Antonini09}).  (4) The orientation
of the stellar binary orbital plane is chosen to be uniformly distributed in
$\cos\phi$ for $\phi\in [0,\pi]$.

For the outer binary, the initial conditions are set as follows. (1) The
orientation of the outer binary is set to satisfy a Gaussian distribution
around the central planes of the stellar disks that initially hosted the
stellar binaries with a standard deviation of $\sim7\arcdeg$$-$$13\arcdeg$,
\citep[c.f.,][]{LuJ09,Bartko09a}. The orientations of the host disks are assumed
to be the same as that of the two best-fit planes, i.e.,
($l$,$b$)=($311\arcdeg$,$-14\arcdeg$) and ($176\arcdeg$, $-53\arcdeg$), respectively
\citep{Luetal10}. (2) The stellar binaries are initially set on orbits with
semimajor axes of $a_{\rm out,i}\sim$0.04$-$0.5$\pc$, and this range is adopted
according to the observational extents of the CWS disk. We note here that the
extents of the structure associated with the second plane are not clear.
Nevertheless, we assume that it is in the same range as the CWS disk. Adopting a
slightly larger range (say, from $0.04\pc$ to $1\pc$)  does not affect the results presented in this section and
next section. (3) The distribution of $a_{\rm out,i}$ of these 
binaries is assumed to follow the same surface density distribution as that of
stars in the CWS disk $f(a_{\rm out,i}) \sim a_{\rm out,i}^{-2.3}$
\citep{LuJ09,Bartko09a}. (4) The distribution of the initial periapsis of the
stellar binaries should depend on detailed mechanisms leading to the
injection. It is not clear how bound binaries (in the outer disk region) are
delivered to the immediate vicinity of the central MBH.  Nevertheless, the
mechanism responsible for the small orbital angular momentum of the stellar
binary may fall into the two extreme categories discussed below.

\begin{itemize}

\item Large perturbations on the orbital angular momenta of binaries
initially rotating around the MBH with insignificant eccentricities: in this
scenario, changes of the orbital specific angular momenta of outer binaries are
large compared with the specific angular momentum required to approach the
tidal radius $r_{\rm tb}$ (denoted by $J_{\rm max}\simeq\sqrt{2G\bh r_{\rm
tb}}$), which leads to an even distribution of the distance to the periapsis of
the outer binaries, i.e., $f(r_{\rm p}) \propto {\rm constant}$. Therefore,
many binaries can penetrate into the vicinity of the MBH with small $D$, and
some of them are broken up during the first close passage but others may be
broken up or merged after many revolutions. Hereafter the model adopting
$f(r_{\rm p})\propto {\rm constant}$ for initially bound binaries with the
above settings is denoted by the ``LP'' model.  The distribution of $r_{\rm p}$
above is similar to that adopted for initially unbound stellar binaries which
is discussed in \citet{Sesana07}.  For comparison, we also introduce a model
(denoted by ``UB'' below) in which the stellar binaries are initially unbound
to the MBH, similar to that used in \citet{Sesana07}. 

\item The perturbations on the orbital specific angular momenta of binary
stars are small compared with $J_{\rm max}$, and $f(r_{\rm p})$ is
significantly different from that for the ``LP'' model. In this scenario, the
eccentricity of an outer binary (or $r_{\rm p}$) may increase (or decrease)
slowly. For example, \citet{MLH09} demonstrated that a secular instability of
eccentric stellar disks rotating around the MBH can gradually excite some
(binary) stars to extremely eccentric orbits ($e\rightarrow 1$). The time
period for this secular evolution can be hundred to thousand times of the
orbital period of the outer binary before the breakup or merging of the stellar
binary. Therefore, the probability for the stellar binary to be broken up is
substantially even at a large $D$ due to the cumulative tidal effect from the MBH
before the binary can possibly migrate into the region with a substantially
small $D$ (as also illustrated in Figure~\ref{fig:f6}). Note that the ejection
velocity of the component escaping away from the MBH is substantially small if
the binary is broken up at a large $D$ rather than at a small $D$. For these
cases, the velocity distribution of the resulted HVSs may be substantially
steeper than that from the LP model simply due to the suppression of the number
of stellar binaries with small $D$. The detailed distribution of $f(r_{\rm
p})$ should depend on the detailed perturbations. Below we adopt a toy model
for this scenario to obtain the spatial and velocity distributions of HVSs and
compare them with that obtained from the ``LP'' model. 

The orbit of the outer binary may be scattered by weak encounters of the binary
star with other stars (in or out of the stellar disk), and its angular momentum
$|J|$ may also be changed after each revolution. We approximate this as a
one-dimensional diffusion process of $|J|$, and the standard deviation of the
change of the angular momentum after each revolution $\sqrt{\langle(\delta
|J|)^2\rangle}={\rm constant}\ll |J|$. This random walk in $|J|$ corresponds to
a change in $r_{\rm p}$ (i.e., $\delta r_{\rm p}=\pm \xi \sqrt{r_{\rm p}/r_{\rm
p,0}}$) after each revolution.  The $r_{\rm p,0}$ should be set by the
penetration parameter (e.g., $D=250$) where the tidal effect begins to be
important for a stellar binary with the maximum semimajor axis considered in
this paper ($a_{\rm b,i}=2\AU$). The periapsis of some stellar binaries may
then diffuse inward to the vicinity of the MBH slowly. During this diffusion
process, these stellar binaries may be broken up at a large $D$ after many
revolutions and few stellar binaries with small $a_{\rm b,i}$ can move into
the very stellar region with small $D$. We denote this model as the ``RW'' model.
For illustration, we set the model parameter $\xi=0.4\AU$ and $r_{\rm p,0}=
(D/100)r_{\rm tb}\sim 630\AU$ with $a_{\rm b}=2\AU$.  A model with a much
larger $\xi$ will be reduced to the ``LP'' model above.  \end{itemize}

Table~\ref{tab:tab2} lists the above models and the settings of a few related
parameters.  According to the models (i.e., the ``UB'' model, the ``LP'' model, and the
``RW'' model), we use Monte Carlo simulations to obtain both the
$\Theta$CDF and the $v$CDF. For each model, the total number of the three-body
experiments is $10^4$.  We only record those cases in which the masses of ejected
stars are in the mass range ($3\msun, 4\msun$) of the detected HVSs if not specified, and then
calculate both the distribution of the inclination angles with respect to the
central planes of their parent population and the velocity distributions.

\begin{figure}
\epsscale{0.90}
\plotone{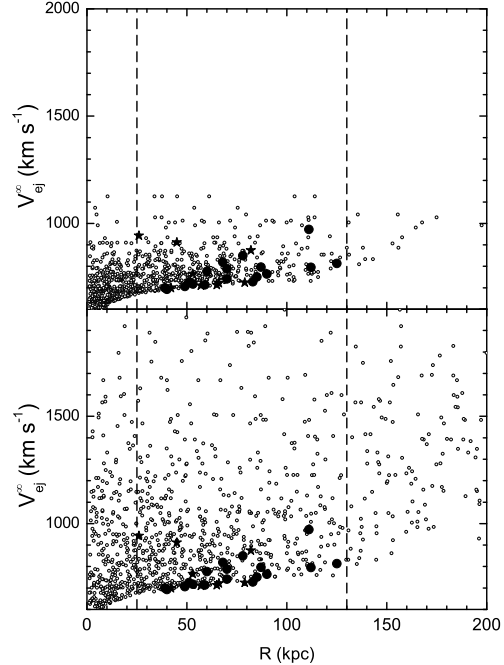}
\caption{Distribution of simulated HVSs in the $v^{\infty}_{\rm ej}-R$
plane ($R$ is the Galactocentric distance of an HVS).  The top and the bottom panels
show the results obtained from the ``RW1'' model and the ``LP1'' model, respectively.
The filled circles and stars represent the detected HVSs that are spatially
associated with the best-fit plane with an orientation almost the same as that
of the CWS disk and the best-fit plane with an orientation similar to that of
the northern arm of the minispiral in the GC (or the outer warped part of the
CWS disk), respectively. The dots represent the simulated HVSs. The apparent
lower boundary is due to the selection effects of HVS candidates in
observations (i.e., the cutoff at the radial velocity $v_{\rm rf}=275 \kms$).
The vertical dashed lines are the lower and upper boundary of the distance of
the detected HVSs from the GC. The total number of the simulated HVSs shown
in the top and the bottom panel are 606 and 449, respectively.
This figure shows that the ejection of HVSs with velocities $>1000 \kms$ is
significantly suppressed in the ``RW1'' model.} \label{fig:f10} \end{figure}

\begin{figure} 
\epsscale{1.0}
\plotone{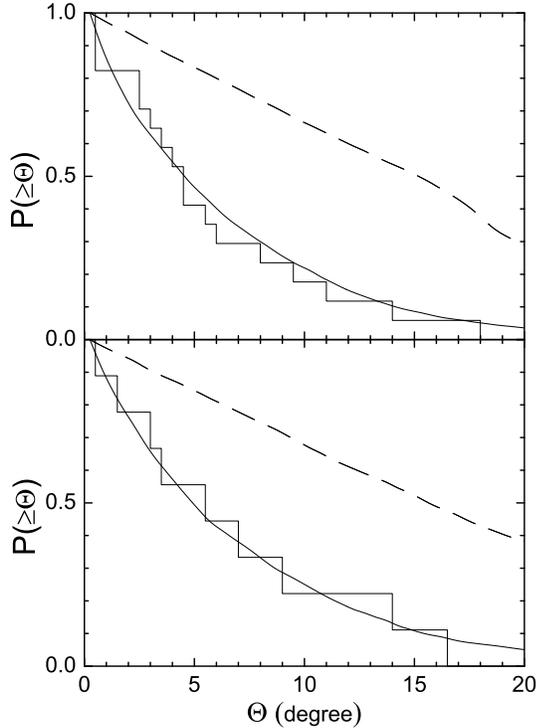} \caption{Cumulative distributions of HVS inclination
angles ($\Theta$CDF) relative to the two best-fit planes of the detected HVSs.
The histograms in the top and bottom panels represent the $\Theta$CDF of the
first and second populations of the detected unbound HVSs, respectively.  The
solid and dashed curves represent the numerical results from the LP-1 model by
adopting the thickness of the two disks which the HVSs are originated from to be
$12\arcdeg$ and $13\arcdeg$, respectively (also see Figure~\ref{fig:f12}). The
dashed curves show the $\Theta$CDFs if the stellar binaries are isotropically
distributed rather than originated from two disks defined by the two best-fit
planes (see details in Section~\ref{sec:spadis}).  } \label{fig:f11} \end{figure}

\begin{figure} 
\epsscale{1.0}
\plotone{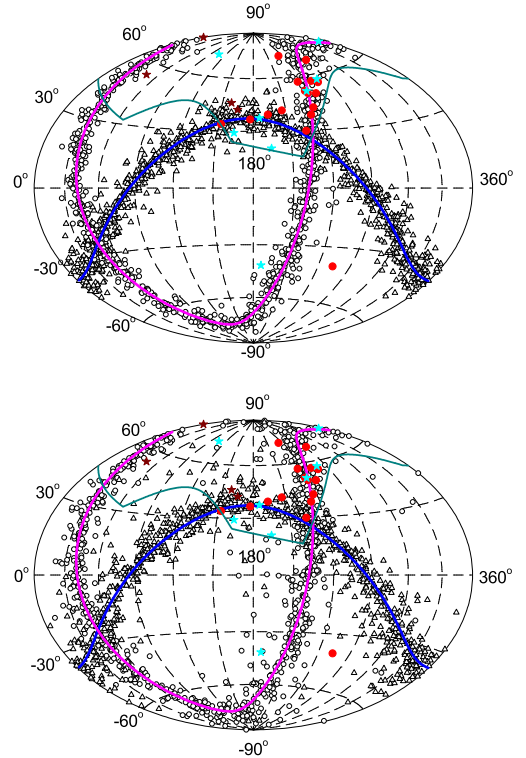} 
\caption{Aitoff-projection for both the simulated HVSs and
the detected HVSs. The red solid circles, the carmine stars, and the cyan stars
represent the detected unbound HVSs, the HVS candidates, and the bound
population of HVSs, respectively. The top and bottom panels show the
simulated HVSs produced by the TBK mechanism and the BBH mechanism (the LP-1
model), respectively. In the top panel, the open circles and triangles
represent the simulated HVSs which are originated from the two best-fit disk
planes with normals of ($l$,$b$)=($311\arcdeg$, $-14\arcdeg$) and ($176\arcdeg$,
$-53\arcdeg$), and the thickness of these two originating disks are $12\arcdeg$ and
$13\arcdeg$, respectively; while in the bottom panel, the thickness of the two
originating disks are $7\arcdeg$ and $10\arcdeg$, respectively. The green solid 
curve shows the boundary of the survey area in the Northern
hemisphere in each panel. This figure shows that the
spatial distribution of the detected HVSs can be well reproduced by both the
TBK mechanism and the BBH mechanism if the progenitors of the detected HVSs are
originated from two thin disks with the same orientations as the fitted ones.
One difference in the resulted spatial distributions of these two mechanisms is
that a small number of simulated HVSs from the BBH mechanism can have larger
inclination angles relative to their original planes, which is due to that some
progenitors can interact with the secondary BH at a very small separation and
receive a large kick. 
} 
\label{fig:f12}
\end{figure}

To compare with the observations summarized in Section~\ref{sec:obs}, the
selection effects must be carefully considered. As shown in
Figure~\ref{fig:f10}, most HVSs are detected in Galactocentric distances from
$25\kpc$ to $130\kpc$ which is partly due to the detection limit and partly due
to the limit in the survey area. In the MMT survey of HVSs by \citet{Brown09a},
the HVS candidates are selected by a cutoff in the radial velocity, i.e.,
$v_{\rm rf}\ga275 \kms$. To account for these selection effects, we
adopt the Galactic potential model listed in Section~\ref{sec:obs} and
simulate the radial distribution of HVSs under the assumption of a constant HVS
ejection rate for each of the above models, which appears to be compatible with
current observations \citep{Brown07}. We only take those HVSs in the surveyed
area with the
radial distance from $25\kpc$ to $130\kpc$ and radial velocities $v_{\rm
rf}\geq275 \kms$ as the simulated sample, which are then used to compare with
the observations. The effect of limited lifetime of the ejected HVSs is also
considered in our calculations. The selection effects are similarly considered
for those models in the BBH mechanism in Section~\ref{sec:BBH} below.
We note here that the
$\Theta$CDF is not affected much but the $v$CDF may be significantly affected
by the selection effects.

\subsection{$\Theta${\rm CDF}: the inclination angle distribution of the 
ejected stars}\label{sec:spadis}

Figure~\ref{fig:f11} shows the $\Theta$CDFs obtained from the ``RW-1'' model. As
seen from this figure, the observational $\Theta$CDF for both HVS populations
can be well reproduced if the HVSs were originated from two disk-like stellar
structures with orientations the same as the two best-fit planes and thickness
of $\sim12\arcdeg$ and $13\arcdeg$, respectively (also see the top panel of
Figure~\ref{fig:f12}). Here, the thickness of a stellar disk is defined by the
standard deviation of the inclination angle of stars in the disk from the disk
central plane.
For these two populations of the detected HVSs, our K-S
tests find the likelihoods of 0.896 and 0.999 that the observational
$\Theta$CDF are drawn from the same distribution as that obtained from the
numerical simulations for the ``RW-1'' model, respectively. All the other models
(i.e., ``LP-1'', ``LP-2'', ``UB-1'', ``UB-2'', and ``RW-2'') can reproduce the observational
$\Theta$CDFs for both HVS populations by choosing suitable thickness (typically
in the range of $7\arcdeg$$-$$13\arcdeg$) for the two disks.  The resulted
$\Theta$CDFs are mainly determined by the thickness of the disks where the HVS
progenitors are originated, and also weakly depend on the mechanism adopted 
that leads to the injection of stellar binaries into the immediate vicinity
of the MBH.

Alternatively, assuming that the stellar binaries, i.e., the HVS progenitors,
were isotropically distributed but with other initial settings the same as
those in the above models, the ejected HVSs should also be isotropically
distributed in the survey area. We do similar numerical simulations and obtain
the $\Theta$CDFs for those isotropically distributed HVSs, relative to the two
best-fit thin disk planes of the detected HVSs, as shown by the dashed curves in Figure~\ref{fig:f11}. Here, a
simulated HVS is assigned to one of the two populations if it is closer to the
best-fit plane of that population than to the other plane. Using the K-S test,
we find the likelihoods of $1\times 10^{-4}$ and $0.016$ that the $\Theta$CDFs
of the simulated HVSs are the same as those from observations for the first and
second population of the detect HVSs, respectively. Therefore, we conclude that
the detected HVSs are highly unlikely to be produced from the tidal breakup of
isotropically distributed progenitorial binary stars. This further 
strengthens the conclusion obtained in \citet{Luetal10}, i.e., the detected HVSs
are probably originated from two thin disks with orientations similar to the
CWS disk and the northern arm of the minispiral (or the warped outer part of
the CWS disk) in the GC, respectively.

\subsection{$v${\rm CDF}: the velocity distribution of the ejected stars}\label{subsubsec:velTBK}

\begin{figure} 
\epsscale{1.0}
\plotone{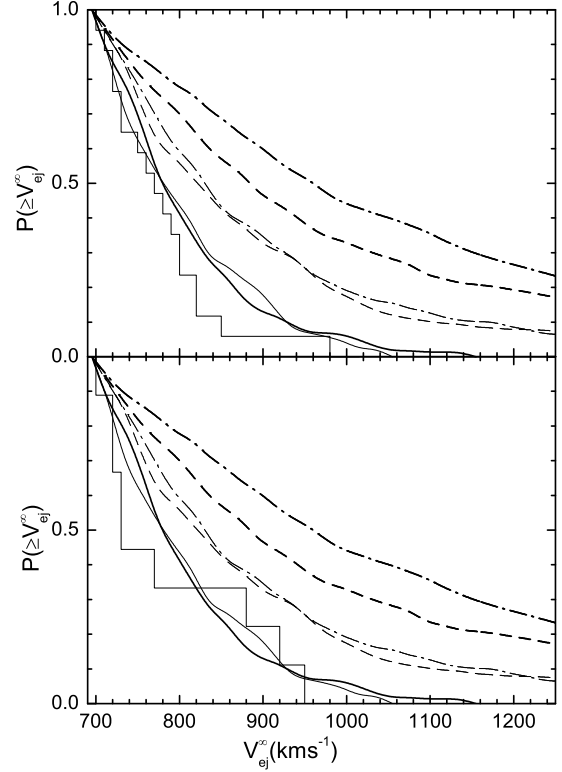} \caption{Cumulative distribution of the ejection
velocity at infinity ($v$CDF) obtained from different models for the TBK
mechanism. The top panel and the bottom panel represent the first and second
populations of the detected HVSs, respectively. The thick (thin) dot-dashed,
dashed, and solid curves represent the $v$CDF obtained from the ``LP-1'' (``LP-2'')
model, the ``UB-1'' (``UB-2'') model, and the ``RW-1'' (``RW-2'') model, respectively.  Both
the ``RW-1'' and ``RW-2'' models can reproduce the observational $v$CDF, while the
other models generate a significant number of HVSs with large $v^{\infty}_{\rm
ej}$ that are not detected by current HVS survey.  } \label{fig:f13}
\end{figure}

Figure~\ref{fig:f13} shows the $v$CDFs for both the simulated HVSs (obtained
from different models) and the observations. Our numerical simulations show
that the $v$CDF is almost independent of the thickness of the disk(s) where the
HVS progenitors are originated, but it does depend on how close the
stellar binaries can approach the MBH and on the initial distribution of
the semimajor axes of the stellar binaries. Different models produce quite
different $v$CDFs. In the ``LP'' models and the ``UB'' models, for example, more than 30\%
of the resulted HVSs (with $v^{\infty}_{\rm ej}\ga700 \kms$) have velocities larger
than the maximum velocity of the detected HVSs ($\sim1000 \kms$), while the ``RW'' models
can produce a steep $v$CDF which is quite similar to the observational ones.
The models with log-normal distributions of $a_{\rm b,i}$ produce less HVSs at
the high-velocity end because of the fraction of stellar binaries with small $a_{\rm b,i}$
(i.e., $\la 0.3\AU$) is substantially smaller compared with that in those models
with the \"{O}pik law.  For the first HVS population, our K-S tests find
$2.7\times 10^{-6}$ ($7.9\times 10^{-3}$) and $2.5\times 10^{-4}$ ($1.3\times
10^{-2}$) likelihoods that the $v$CDFs obtained from the ``LP-1'' (``UB-1'') model and
the ``LP-2'' (``UB-2'') model are drawn from the same distribution as the observational
ones, respectively, which suggests that the first HVS population is unlikely to
be produced by either of the ``LP'' model and the ``UB'' model. For the second HVS
population, the K-S likelihood is $0.01$ ($0.05$) and $0.06$ ($0.07$) for the
``LP-1'' (``UB-1'') model and the ``LP-2'' (``UB-2'') model, respectively. These numbers
suggest that the second population is not likely to be produced by the ``LP'' or ``UB''
models with an initial $a_{\rm b,i}$ distribution of $1/a_{\rm b,i}$ but it may
not be inconsistent with the ``LP'' (or ``UB'') models with a log-normal distribution of
$a_{\rm b,i}$ (though with limited statistics).  However, the $v$CDFs resulted from
the ``RW'' models appear to be consistent with the observations as the K-S tests
find the likelihoods of 0.52 (0.13) and 0.43 (0.52) that the observational
$v$CDFs of the first (second) HVS population are the same as that obtained from
the ``RW-1'' model and the ``RW-2'' model, respectively.

Adopting a different form of the Galactic potential may affect the estimation of the
$v$CDF for the detected HVSs in Section 2 and the selection effects discussed
in Section 4.1. For example, if adopting a simple Galactic potential model as
that described by Equation (8) in \citet{Kenyon08}, the $v^{\infty}_{\rm ej}$
of the detected HVSs ranges from $850 \kms$ to $1200 \kms$ and the slope of the
$v$CDF is slightly flatter than that estimated in Section 2. The simulated
$v$CDFs from the models of ``LP-1'', ``LP-2'', ``UB-1'', and ``UB-2'' are not likely to be 
consistent with the
$v$CDF of the detected HVSs, while both the ``RW-1'' model and the ``RW-2'' model can
produce a $v$CDF similar to that estimated for the detected HVSs according to
the new Galactic potential. 
Our main conclusion that the TBK mechanism can reproduce the detected $v$CDF
made in this
section is not affected by the choice of the
Galactic potential (also see discussion in \citet{Sesana07} and \citet{Kenyon08}).

To close this section, we note here that the fraction of stellar binaries 
resulting in ejection of HVSs with properties similar to the detected ones
is around $\sim3\%$$-$$6\%$ in those models adopted above (see Table~\ref{tab:tab2}). Current
observations imply that the total number of HVSs similar to the detected ones
is $\sim100$ \citep{Brown07}. Therefore, the number of stellar binaries
is required to be around a few thousands and the injecting rate is $\sim10^{-5}{\rm yr}^{-1}$. This rate appears to be roughly consistent with that 
estimated by \citet{MLH09} if the progenitors are injected into the immediate 
vicinity of the MBH due to secular instability developed in the stellar disk.
However, we caution here that the dynamics leading to the injection of (binary)
stars is currently not clear, and therefore any argument based on the event 
rate may have substantial limitation in distinguishing the production mechanism
of the detected HVSs.

\section{Ejecting HVSs by a hypothetical BBH in the GC}\label{sec:BBH}

The dynamical interactions between a BBH and single stars can also eject HVSs
(the BBH mechanism). In Section~\ref{subsec:paraspace}, we discuss possible
constraints on the possible parameter space of a BBH if it exists (or existed)
in the GC and is responsible for the ejection of the detected HVSs. According
to those constraints, in Sections~\ref{subsec:paraspace}--\ref{subsec:BBHvCDF},
we perform a large number of Monte Carlo simulations for a few models of the
decay of the BBH orbit to investigate whether the spatial and velocity
distributions of the detected HVSs are compatible with the BBH mechanism.\footnote{If multiple intermediate-mass BHs exist in the GC \citep{Zwart06},
HVSs can also be ejected from the GC by interactions with the BHs; but the
stability of the system is not clear and an {\it N}-body simulation of ejection of
HVSs from such a system is beyond the scope of this paper.}

\subsection{Parameter space for the hypothetical BBH}\label{subsec:paraspace}

\begin{figure} 
\epsscale{1.0} 
\plotone{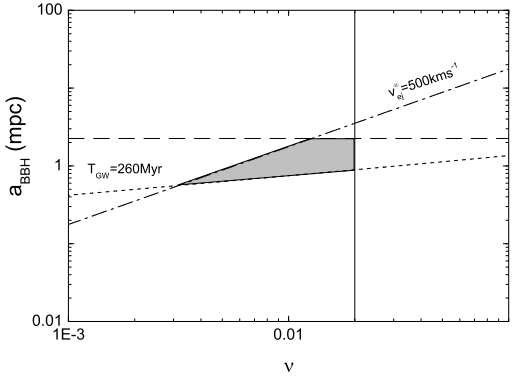} \caption{Parameter space for
the hypothetical BBH in the GC if it is responsible for the ejection of the
detected HVSs. The short dashed line represents the upper limit of the current
semimajor axis of the BBH that is compatible with other observations
\citep{GM09,YT03}. The vertical solid line represents the upper limit of the
mass ratio of the BBH if the secondary BH now exists in the GC. The dot-dashed
line is the constraint on the initial semimajor axis of the BBH from the HVS
ejection velocity (see Section~\ref{subsec:paraspace}). The dotted line
is the constraint on the initial semimajor axis of the BBH from the longest
travel time of the detected HVSs, which should be smaller than the decay
timescale of the BBH only by gravitational wave radiation. See details in
Section~\ref{subsec:paraspace}.} \label{fig:f14} \end{figure}

The parameter space for the secondary BH, if being currently located close to the central
primary MBH, has been investigated by \citet{HM03}, \citet{YT03},
\citet{Gillessenetal09}, and \citet{GM09}. The observations so far have not
shown evidence in contradict with the existence of a BBH with semimajor
axis $a_{\rm BBH}\la 2\mpc$ and mass ratio $\nu\equiv\bhs/\bhsp <0.02$
(see Figure~2 in \citet{YT03} and Figure~13 in \citet{GM09}), where $\bhsp$ and $\bhs$ are the masses of
the two components of the BBH. If a BBH is responsible for ejecting the
detected HVSs, simple constraints on the BBH parameter space may be directly
obtained by the properties of the HVSs as follows (see Figure~\ref{fig:f14}).

\begin{itemize}

\item The travel time of the detected HVSs ranges from $30$ to $260$~Myr so that
the BBH should exist in the GC $260$~Myr ago and should not have merged into a
single BH $30$~Myr ago, which put a stringent constraint on the location of the
hypothetical BBH in the GC. The timescale for the decay of the orbit of a BBH
in the gravitational wave radiation dominated stage is given by \citep{Peters64}
\begin{eqnarray}
T_{\rm GW} & = & \frac{a_{\rm BBH}}{|\dot{a}_{\rm BBH}|} \nonumber \\
           & = & \frac{5}{64}\frac{c^5}{G^3}
             \frac{a_{\rm BBH}^4}{\bhsp\bhs(\bhsp+\bhs)}f^{-1}(e) \nonumber \\
           & = & 3.6\times10^7{\rm yr} \frac{f^{-1}(e)}{\nu+\nu^2}
             \left(\frac{a_{\rm BBH}}{\mpc}\right)^4 \left(\frac{\bhsp}{4\times
            10^6\msun}\right)^{-3}, \nonumber \\
\label{eq:TGW}
\end{eqnarray}
where $\dot{a}_{\rm BBH}=da_{\rm BBH}/dt$ and
\be
f(e)\equiv (1-e^2)^{-7/2}\left(1+\frac{73}{24}e^2+\frac{37}{96}e^4\right).
\ee
Combining Equation (\ref{eq:TGW}) and the constraint from the HVS travel time,
the hypothetical BBH should have a semimajor axis $a_{\rm BBH,0}\ga
2.2\left(\nu+\nu^2 \right)^{1/4} \mpc$ when the detected HVSs began to be
ejected out from the GC (see the long-dashed line in Figure~\ref{fig:f14}).
Note that the above constraint on $a_{\rm BBH,0}$ is obtained by assuming that
the initial eccentricity of the BBH is $e_{\rm BBH,0}=0$ ($f(e_{\rm BBH,0})=1$)
and the BBH orbital decay is dominated by the energy loss due to gravitational
wave radiation \citep[for details see][]{Peters64}. Adopting a moderately large
initial BBH eccentricity does not change the constraint much (e.g., $a_{\rm
BBH,0}$ is increased only by a factor of 2.5 if $e_{\rm BBH,0}=0.5$).  However,
the constraint can be changed by orders of magnitude if the initial
eccentricity is close to 1. 

\item Before the gravitational wave radiation dominated stage, the orbital
decay of the BBH when it is ``hard'' with $a_{\rm BBH,0}\la
G\bhs/4\sigma^2\simeq 430\nu(\sigma/100 \kms)^{-2}\mpc$ \citep{Quinlan96} should
be dominated by the interactions of the BBH with unbound stars in the Galactic
bulge or bound stars in the GC stellar cusp \citep[e.g.,][]{Yu02,Sesana08}. And
the interactions between the BBH and the stars injected from the stellar disk
(e.g., the CWS disk) may be insignificant to the BBH orbital decay as the
number of those stars injected into the vicinity of the BBH is relatively small
compared with that from other bound stars. Although HVSs can be ejected out as
the high-velocity tail when the BBH semimajor axis is large (e.g., from
$0.2\pc$ to a few $\mpc$, see \citealt{Sesana08} and \citealt{Baumgardt06}),
the ejecting efficiency is low and the majority of the ejected high-velocity
stars should be old stars from the cusp rather than young stars originated from
the stellar disk. If the progenitors of the detected HVSs were originated from
the stellar disk surrounding the MBH and the HVSs are ejected out by the BBH
mechanism, the BBH (with a given mass ratio) should be hard enough to
efficiently eject such high-velocity stars. 

The rms velocity of the ejected HVSs at infinity, if their progenitors are on
parabolic (or hyperbolic) orbits with eccentricities close to 1, can be given
by \citep{YT03}
\begin{equation}
\sim 740 \kms \left(\frac{\nu}{0.01}\right)^{1/2}
\left(\frac{a_{\rm BBH}}{\mpc}\right)^{-1/2}
\left(\frac{\bhsp}{4\times10^6\msun}\right)^{1/2}.
\label{eq:vel2}
\end{equation}
Therefore we should have a rough constraint on the BBH semimajor axis when the
BBH began to eject those detected HVSs, i.e., $a_{\rm BBH,0}\la110 \nu\mpc$,
as the minimum velocity of these HVSs at infinity is $\sim700 \kms$ (the solid
line in Figure~\ref{fig:f14}). Considering that HVSs can be ejected by the BBH
with a larger semimajor axis, we reset $a_{\rm BBH,0}\la 220\nu\mpc$ which
corresponds to an rms ejection velocity of $\sim500 \kms$ and is substantially
smaller than the value at which the BBH becomes ``hard'' \citep{Quinlan96}.

\end{itemize}

\subsection{Models for the orbital evolution of the BBH}\label{sec:BBHmodel}

The orbital evolution of a BBH during its hard stage is dominated by three-body
interactions of low-angular momentum stars with the BBH. 
As shown in \citet{YT03}, the decay timescale for the hypothetical BBH in the
GC is roughly 4--8~Gyr during the hard stage unless the potential of the
Galactic bulge is significantly flattened or triaxial \citep{Yu02} or there are
a large number of massive perturbers \citep{Perets07} which lead to efficient
transferring of stars onto low-angular momentum orbits. This decay timescale is
substantially longer than the travel time of the detected HVSs ($\leq
260$~Myr). According to these estimates and the constraints on the parameter
space of the BBH, we can use the following three simple models to generally describe
the orbital evolution of the BBH during the course of ejecting the detected
HVSs. A few parameters involved in these models are listed in
Table~\ref{tab:tab3}.

\begin{description}

\item{(a)} If the potential of the Galactic bulge is spherical or at least is
not significantly flattened/triaxial and the massive perturbers in the Galactic
bulge is not sufficient in quickly filling the loss-cone, then the orbital
decay of the BBH is slow. In this model, we assume that the semimajor axis of
the BBH does not change during the period of ejecting the detected HVSs and
$a_{\rm BBH,0}\simeq 220\nu\mpc$. (Choosing a somewhat smaller value does not
affect our results.)

\item{(b)} If the BBH is initially at $a_{\rm BBH,0}=2.2\left(\nu+\nu^2
\right)^{1/4}\mpc$ and the orbital decay is dominated by gravitational wave
radiation, we have the evolution of the semimajor axis of an initially circular
BBH as $a_{\rm BBH}=a_{\rm BBH,0}(1-t/{260\rm Myr})^{1/4}$. For a BBH with a
moderately large initial eccentricity $e_{\rm BBH,0}$, the orbital evolution is
similar to the circular one. For a BBH with extremely large initial $e_{\rm
BBH,0}$ \citep[e.g., see][]{Makino07}, the decay timescale is too short and
thus the BBH should not be able to eject HVSs over a time span similar to the
longest travel time of the detected HVSs.

\item{(c)} If the orbital decay of the BBH during the hard stage is much faster
than that of the spherical case because of a highly flattened or triaxial bulge
\citep{Yu02} or many massive perturbers in the bulge \citep{Perets07}, the
semimajor axis of the BBH may quickly evolve from the upper boundary to the
lower boundary shown in Figure~\ref{fig:f14}. In order to account for the
detected HVSs, this decay timescale should be comparable to the span of the HVS
travel time.  Otherwise, it is either reduced to model (b) or simply
inconsistent with observations. In this model, we assume an exponential
evolution for the BBH semimajor axis, i.e., $a_{\rm BBH}=a_{\rm
BBH,0}\exp(-t/260{\rm Myr}) \simeq 220\nu\exp(-t/260{\rm Myr})\mpc$, which
corresponds to a constant $a_{\rm BBH}/|\dot{a}_{\rm BBH}|$.  

\end{description}

With the above models, we perform Monte-Carlo simulations of the three-body
interactions between a single star and the hypothetical BBH with mass ratios
$\nu=0.01$ and 0.003, respectively (see Table~\ref{tab:tab3}). For each
model, the total number of the three-body experiments is $10^4$. We set the 
tolerance of the fractional energy error to be $10^{-9}$ for each three-body
experiment.
The initial
conditions for the injecting stars are set to be similar to those given in
Section~\ref{sec:Result} for the outer binaries. The initial distribution of
the penetration parameters, now defined as $r_{\rm p}/a_{\rm BBH,0}$, is set to
be $f(r_{\rm p}/a_{\rm BBH,0})\propto {\rm constant}$ (similar to the ``LP'' models
for the TBK mechanism), and the starting point is $r_{\rm p,0}/a_{\rm
BBH,0}\sim$1.1$-$1.2 because the ejection efficiency and velocity drop rapidly
for $r_{\rm p}>r_{\rm p,0}$ \citep[e.g.,][]{Sesana06}.  The orientation of the
BBH relative to the normal of the stellar disk(s) is set to a few typical
values, e.g., $\theta=0$, $\pi/4$, $\pi/2$, etc., and our calculations show the
resulted $\Theta$CDF and $v$CDF are slightly affected but not much by the choice
of this orientation. We can also choose the ``RW'' models for the BBH mechanism
(similar to that for the TBK mechanism), but our calculations show this model
do not lead to a suppression of the ejection of HVSs at the high-velocity end.

\begin{deluxetable}{lclccc}
\tablecaption{Different models for the BBH mechanism}
\tablehead{
\colhead{$a_{\rm BBH,0}$} & \colhead{$\nu$} &
\colhead{Decay Model} & \colhead{$\theta$} & {\rm Origin}$^1$ & \colhead{Notations} }
\startdata
2.20\mpc  & 0.01  & (a) Fixed $a_{\rm BBH}$  & $\pi/4$ & {\rm Disk} & FIX   \\
0.55\mpc  & 0.003 & (b) GW dominant & $\pi/4$ & {\rm Disk} & GW1  \\
0.75\mpc  & 0.01  & (b) GW dominant & $\pi/4$ & {\rm Disk} & GW2  \\
0.75\mpc  & 0.01  & (b) GW dominant & $\pi/2$ & {\rm Disk} & GW3  \\
2.20\mpc  & 0.01  & (c) Exponential decay    & $\pi/4$ & {\rm Disk} & ExpD   \\
2.20\mpc  & 0.01  & (a) Fixed $a_{\rm BBH}$  & $\pi/4$ & {\rm Infinity} & FIX-UB 
\enddata
\tablenotetext{1}{The origin of the injecting stars is set to be from
a disk similar to the CWS disk for the first five  models listed here, while 
it is set to be from infinity and unbound to the BBH for the last model (the
``FIX-UB'' model).}
\label{tab:tab3}
\end{deluxetable}

\subsection{$\Theta${\rm CDF}}\label{subsec:BBHthetaCDF}

\begin{figure} 
\epsscale{1.0}
\plotone{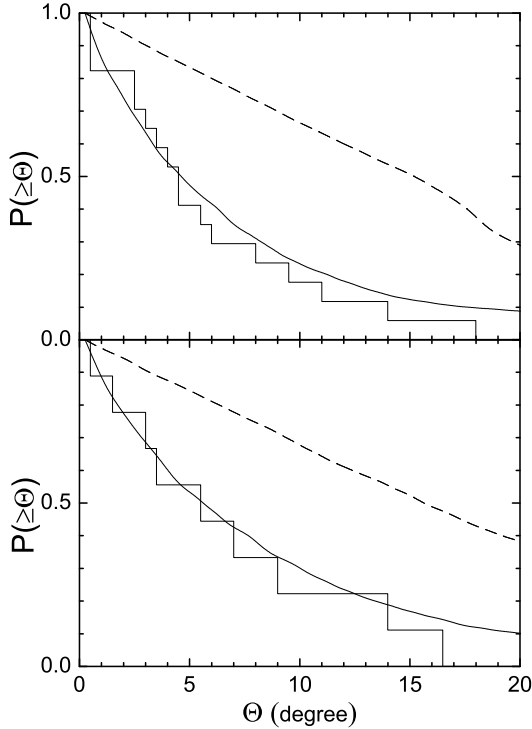} \caption{
Legends are the same as in Figure~\ref{fig:f10}, but the solid curves represent
the $\Theta$CDF obtained from the ``GW2'' model in the BBH mechanism. The solid
curves in the top and bottom panels show the $\Theta$CDF of the simulated HVSs
which are originated from the two disk-like structures with orientations the
same as the two best-fit planes and thickness of $7\arcdeg$ and $10\arcdeg$,
respectively.
} \label{fig:f15} \end{figure}                                                  

Figure~\ref{fig:f15} shows the $\Theta$CDFs obtained from the ``GW2'' model listed
in Table~\ref{tab:tab3} for the BBH mechanism. As seen from the figure, the
observational $\Theta$CDFs can be well reproduced by this model if the HVS
progenitors are originated from the two best-fit disks with thickness of
$7\arcdeg$ and $10\arcdeg$, respectively (also see the Aitoff-projection of the
simulated HVSs in the bottom panel of Figure~\ref{fig:f12}).  For these two
populations of the detected HVSs, our K-S tests find the likelihoods of 0.97
and 0.91 that the observational $\Theta$CDF are drawn from the same
distributions as that obtained from the numerical simulations for the ``GW2'' model,
respectively.  All the other models listed in Table~\ref{tab:tab3} can produce
$\Theta$CDFs similar to the observational ones by choosing suitable thickness
of the two disks (typically in the range $7\arcdeg$$-$$10\arcdeg$) where the HVS
progenitors are originated. As seen from Figure~\ref{fig:f12}, a number of HVSs
resulted from the BBH mechanism can have substantially large inclination angles
relative to their original planes; while no such HVSs appears in the TBK
mechanism. The reason for this difference is that there always exist some HVS
progenitors which can interact with the secondary BH at a very small separation
and receive a relatively large kick.  These very close interactions lead to the
large inclination angles of some simulated HVSs. This difference may be helpful
in distinguishing the HVS production mechanism in future.

The deviation of the simulated HVSs from the central plane of
the disk(s) (with a given thickness) is larger if the mass ratio of the BBH is
larger (see the analysis in \citealt{Luetal10}). But for the allowed parameter
space of the hypothetical BBH as shown in Figure~\ref{fig:f14}, the differences
in the resulted $\Theta$CDFs are quite small for different choices of the mass
ratio (and the semimajor axis) and they are easily compensated by slightly
different choices of the disk thickness. Similarly, a different choice of the
orientation of the disk plane relative to the orbital plane of the BBH may also
result in slightly different $\Theta$CDF, but which can also be compensated by
different choices of the thickness of the disk(s).

\subsection{$v${\rm CDF}}\label{subsec:BBHvCDF}

\begin{figure} 
\epsscale{1.0}
\plotone{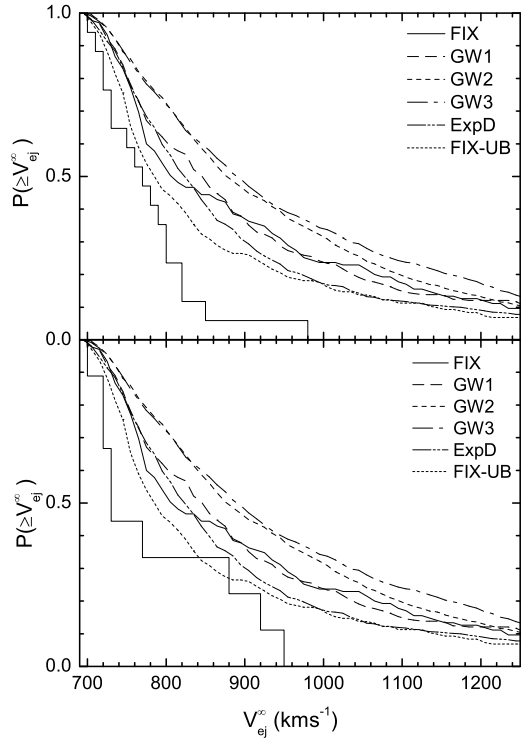} \caption{Cumulative
distribution of the ejection velocity at infinity ($v$CDF) obtained from
different models for the BBH mechanism. The histograms in the top and
bottom panels represent the observational $v$CDFs for the first and second
populations of the detected HVSs, respectively. The solid, long-dashed, 
short-dotted, dot-dashed, dot-dot-dashed, and dotted curves represent the $v$CDF
obtained from the ``FIX'' model, the ``GW1'' model, the ``GW2'' model, the ``GW3'' model,
the ``ExpD'' model, and the ``FIX-UB''
model, respectively.} \label{fig:f16} \end{figure}

Figure~\ref{fig:f16} shows the $v$CDFs for both the detected HVSs and the
simulated HVSs (obtained from different evolutionary models for the BBH 
mechanism listed in Table~\ref{tab:tab3}). Our calculations show that the
$v$CDF (with $v^{\infty}_{\rm ej}>700 \kms$) is almost independent of the 
choice of the thickness of the disk where the HVS progenitors are originated.
The simulated $v$CDF is insensitive to the choice of the orientation of the disk
relative to the BBH orbital plane. The simulated $v$CDF is also insensitive to
the choice of the mass ratio $\nu$ (if $\nu\sim0.01$$-$$0.001$) and the eccentricity of the BBH \citep[also
see][]{Sesana07}. As seen from Figure~\ref{fig:f16}, different models produce
quite similar $v$CDFs and too many HVSs with ejection velocities
substantially higher than those of the detected HVSs. The primary reason for
these similarly flat $v$CDFs is as follows. The HVS progenitors injected into
the region with penetration parameter $r_{\rm p}/a_{\rm BBH}\la 1$ can gain
some energy during the dynamical interaction with the BBH, and the mean energy
gain is mainly determined by the semimajor axis and the mass ratio of the BBH 
\citep{Quinlan96,Sesana06} but insensitive to the detailed values of 
$r_{\rm p}/a_{\rm BBH}$. With increasing $r_{\rm p}/a_{\rm BBH}$ to be slightly
larger than 1, the energy gain dramatically drops to 0 
\citep[also see][]{Sesana06}. Therefore, the slope of the $v$CDF is largely
determined by the scatters of the energy gains around their rms value at
$r_{\rm p}/a_{\rm BBH}\la 1$, and the scatters are mainly due to the different
orbital phases of the BBH and insensitive to other BBH parameters. Because of
these features in the BBH mechanism, even the ``RW'' model for 
injecting stars, which suppresses the ejection of HVSs at the high-velocity end
for the TBK mechanism, does not lead to such a suppression for the BBH mechanism here.

With the relevant parameters for the models listed in Table~\ref{tab:tab3}, 
the simulated $v$CDFs are all quite flat (except for the ``FIX-UB'' model), 
compared with the observational ones. For the first HVS population, our K-S
tests find that the likelihoods for the simulated $v$CDFs to be consistent 
with the observed distribution are $0.01$, $0.18$, $0.003$, $0.0001$, 
$0.0001$, and $0.01$ for the ``FIX'', ``FIX-UB'', ``GW1'', ``GW2'', ``GW3'', 
and ``ExpD'' models, respectively. Except the high likelihood for the 
``FIX-UB'' model, the other small K-S likelihoods above suggest that the first population 
of the detected HVSs are unlikely drawn from the same distributions as those
obtained from the other models.
For the second HVS population, the K-S likelihoods are $0.03$, $0.12$, $0.05$,
$0.02$, $0.02$, and $0.04$ for the ``FIX'' , ``FIX-UB'', ``GW1'', ``GW2'', ``GW3'',
and ``ExpD'' models, respectively. Those likelihoods are not small enough so
that the $v$CDF for the second HVS population does not appear to be inconsistent
with those obtained from the those models, especially for the ``FIX-UB'' model.
For the combined sample of both the first and the second HVS populations, the
KS likelihoods are $0.03$, $0.14$, $0.003$, $3\times 10^{-5}$, $3\times 10^{-5}$,
and $0.02$, respectively.

For the models in which $a_{\rm BBH}$ evolves with 
time (i.e., the ``GW'' and ``ExpD'' models), our simulations indicate a weak
dependence of the ejection velocity on time, as the BBH semimajor axis is 
smaller at a later time and the corresponding ejection velocity is larger 
(see Equation (9)). Compared to our results, the simulated $v$CDF obtained by \citet{Sesana07}
is flatter, as in their model the BBH decays fast at large separations and
more stars with higher velocities are ejected at a later time
as the BBH becomes harder.
Due to the limited statistics of the detected HVSs and the uncertainties in the BBH
dynamical evolution model, it appears to be premature to use the $v$CDF to distinguish
whether the HVSs are ejected due to the TBK mechanism or the BBH mechanism and
also constrain detailed dynamical models for injecting stars to the MBH(s) vicinity.

Note that the ejected stars are actually moving in the Galactic potential,
some of those with low velocities (e.g., $v^{\infty}_{\rm ej}\la400 \kms$) 
may return to the vicinity of the BBH and receive further kicks, which may
change the final $v$CDF. We check this effect by doing additional three-body
experiments for the ejected stars with low ejection velocities. We find that
the obtained final $v$CDFs are only slightly flatter than those shown in 
Figure~\ref{fig:f16} and our conclusions above will not be significantly affected.

\section{Conclusions and discussions}\label{sec:Conclusion}

In this paper, we first use three-body experiments to study the interactions of
the MBH with binary stars bound to the MBH. We find that the probability of
ejecting HVSs is substantially enhanced due to multiple encounters between the
MBH and the stellar binaries injecting into its vicinity even at a distance
substantially larger than the tidal breakup radius. Given a penetration
parameter (e.g., $D\ga 150$), the velocities of the HVSs ejected after multiple
encounters are substantially smaller than those ejected by first encounters
because of the excitations of their semimajor axes and eccentricities.

Assuming that the progenitors of the detected HVSs are originated from stellar
disk structures, by using Monte Carlo simulations we find the distribution of
the inclination of HVSs relative to the disk planes can be well reproduced by
both the mechanism of tidal breakup of binary stars and the mechanism of
ejecting HVSs by a hypothetical BBH in the GC. However, an isotropical
origination of HVS progenitors is inconsistent with the observed inclination
distributions. We find that the spatial distribution of HVSs is primarily
determined by the geometrical structure(s) that their progenitors originated
but less sensitive to whether the ejection is due to tidal breakup of binary
stars in the vicinity of the central MBH or dynamical interactions of stars
with a BBH.  These results strengthen the conclusion in \citet{Luetal10}
that the detected HVSs are probably originated from the two disk-like stellar
structures in the GC, one of which is probably the CWS disk.

We find that the HVS velocity distribution should encode some dynamical
information in the GC environment. Assuming that the detected HVSs were ejected
out from the GC by tidal breakup of binary stars, its velocity distribution can
be reproduced if the HVS progenitors diffuse onto low angular momentum orbits
slowly and most of the progenitorial binaries were broken up at large distances
with small ejection velocities. In this scenario, the HVS velocity distribution
not only depends on the distribution of the initial semimajor axes of the
stellar binaries but also on how the binary stars diffuse onto
low angular momentum orbits to reach the vicinity of the central MBH. If the
progenitors were injected to the vicinity of the MBH by large perturbations on
their orbital angular momenta, the simulated HVS velocity distributions appear
to be flatter at the high-velocity end, which are inconsistent with the
observed ones.  If the HVSs were ejected out by a BBH in the GC, our
simulations show that the HVS velocity distribution is not sensitive to the
mass ratio and other properties of the BBH but may depend on how the BBH orbit
decays (cf., \citealt{Sesana07}).  We find that it is less likely that the observed
HVS velocity distribution can be reproduced with the allowed parameter space of
the BBH, as the BBH mechanism produces a relatively flat spectrum at the
high-velocity end; however, the BBH mechanism cannot be statistically ruled
out, yet. Future deep surveys of HVSs and better statistics of the HVS spatial
and velocity distributions should enable to distinguish the ejection mechanisms
of HVSs and shed new light on the dynamical environment surrounding the central
MBH.

So far, most of the detected HVSs are B-type stars, and their progenitors are
associated to some specific stellar structures, such as the CWS disk in the GC.
However, there are many late-type stars in the stellar cusp of the GC and
injection of these late-type (binary) stars into the vicinity of central MBH
(or BBH) should result in an old population of HVSs in the Galactic halo. The
spatial distribution of the old population of HVSs should be more isotropic as
the parent population of their progenitors is dynamically relaxed. The velocity
distribution of this old-population HVSs could also be quite different, as
their progenitors may have a different dynamical environment and are injected
into the vicinity of the central MBH differently.  Searching for the old
population of HVSs \citep{KG07,Kollmeier09} should be helpful in revealing the
dynamical environment in the GC.

{\bf Note added in proof:} The unbound HVS, HE 0437-5439, was proposed to be ejected
  out from the LMC (Edelmann et al. 2005); and this object was excluded in the
  original fits to the two disk planes in Lu et al. (2010).  However, a recent
  measurement of its proper motion by the HST suggests it should be originated
  from the GC (Brown et al. 2010). As seen from Figure 12, the
  deviation of HE 0437-5439 away from the disk plane(s) is relatively large
  compared with other HVSs, and it is difficult to have its progenitor
  originated from the best fit disk(s) under the TBK mechanism. However, the
  deviation can be compatible with the disk origin under the BBH mechanism in      
  which a small fraction of HVSs with large scatters can be produced (see
  discussion in Section 5.3). Because of the short lifetime and long traveling
  time of HE 0437-5439, it is suggested that this object was originally a
  binary star when being ejected out from the GC and lately merged into a
  single blue straggler (Brown et al. 2010, Perets 2009b). This proposal is also compatible
  with the BBH mechanism in which hypervelocity binary stars can be ejected
  out as predicted in Lu et al. (2007).

\acknowledgements
We thank Dr. Warren Brown for helpful conversations on the MMT HVS survey and
sending us the data on the boundary of the survey area.
This work was supported in part by the BaiRen program from the National 
Astronomical Observatories, Chinese Academy of Sciences, and the National 
Natural Science Foundation of China under nos. 10843009, 10973001, and 10973017.

\end{document}